\documentclass[usegraphicx]{mn2e}
\usepackage{amssymb}
\newif\ifAMStwofonts

\def\ee #1 {\times 10^{#1}}
\def\ut #1 #2 { \, \rmn{#1}^{#2}}
\def\u #1 { \, \rmn{#1}}

\def\persec {\, \hbox{s}^{-1}}
\def\percc {\,\rmn{cm}^{-3}}
\def\persqcm {\, \hbox{cm}^{-2}}
\def\micron {\, \mu \hbox{m}}
\def\half{{\textstyle \frac{1}{2}}}
\def\thalf{{\textstyle{ 3\over 2}}}
\let\grad=\nabla
\def\cross{\bmath{\times}}
\def\curl #1 {\grad \cross #1}
\def\div #1 {\grad \cdot #1}

\def\rau{\left(\frac{r}{1 AU}\right)}
\def\lau{\left(\frac{L_\star}{L_\odot}\right)}
\def\mau{\left(\frac{M_\star}{M_\odot}\right)}
\def\mol{\left(\frac{2.34}{\mu}\right)}

\def\e{\bmath{e}}
\def\v{\bmath{v}}

\def\B{\bmath{B}}
\def\E{\bmath{E}}            
\def\Bh{\bmath{\hat{B}}}
\def\fh{\bmath{\hat{\phi}}}  
\def\rh{\bmath{\hat{r}}}     

\def\vr{v_{r}}
\def\vf{v_{\phi}}            
\def\vk{v_{K}}               

\def\Epa{\bmath{E'_\parallel}}  
\def\Epe{\bmath{E'_\perp}}  
\def\J{\bmath{J}}

\def\dv{\bmath{\delta\v}}
\def\dE{\bmath{\delta\E}}
\def\dB{\bmath{\delta\B}}
\def\dJ{\bmath{\delta\J}}

\newcommand{\delt} [1] {\frac{\partial #1}{\partial t}}

\input epsf

\title{Magnetorotational instability in protoplanetary discs}
\author[R. Salmeron and M. Wardle]
       {Raquel Salmeron$ ^1 $ \& Mark Wardle$ ^2 $ \\
$ ^1 $School of Physics, University of Sydney, NSW 2006, Australia \\
$ ^2 $Physics Department, Macquarie University, NSW 2109, Australia}

\date{2004 August 23}
\pagerange{\pageref{firstpage}--\pageref{lastpage}}
\pubyear{2004}
\begin{document}
\maketitle
\label{firstpage}
\begin{abstract}
We investigate the linear growth and vertical structure of the 
magnetorotational instability (MRI) in weakly ionised, stratified accretion 
discs. The magnetic field is initially 
vertical and dust grains are assumed to have settled towards the midplane, so 
charges are carried by electrons and ions only.
Solutions are obtained at representative radial locations from the central 
protostar for different choices of the initial magnetic field strength, 
sources of ionisation, disc structure and configuration of the conductivity 
tensor. 

The MRI is active over a wide range of magnetic field strengths 
and fluid conditions in low conductivity discs. Moreover, no 
evidence was found of a low-limit field strength below which unstable modes do 
not exist. 
For the minimum-mass solar 
nebula model, incorporating cosmic ray ionisation, perturbations grow at $1$
AU for $B \lesssim 8$ G. For a significant subset of these 
strengths ($200$ mG $\lesssim B \lesssim 5$ G), the growth rate is of order 
the ideal MHD rate ($0.75 \Omega$). Hall 
conductivity modifies the structure and growth rate of global unstable modes 
at $1$ AU for all magnetic field strengths that support MRI. As a result, at 
this radius, modes obtained with a full conductivity 
tensor grow faster and are active over a more extended cross-section of the 
disc, than perturbations in the ambipolar diffusion limit. For relatively 
strong fields (e.g. $B \gtrsim 200$ mG), ambipolar diffusion alters the 
envelope shapes of the unstable modes, which peak at an intermediate 
height, instead of being mostly flat as modes in the Hall limit are in this 
region of parameter space. Similarly,
when cosmic rays are assumed to be excluded from the disc by the winds emitted 
by the magnetically active protostar, unstable modes grow at this radius for 
$B \lesssim 2$ G. For strong fields, perturbations exhibit a kink at the 
height where x-ray ionisation becomes active.

This study shows that, despite the low magnetic coupling, the magnetic field 
is dynamically important for a large range of fluid conditions and field 
strengths in protostellar discs. An example of such magnetic activity is the 
generation of MRI unstable modes, which are supported at $1$ AU for field 
strengths up to a few gauss. Hall diffusion largely determines the 
structure and growth rate of these perturbations for all studied radii. At 
radii of order $1$ AU, in particular, it is crucial to incorporate the full 
conductivity tensor in the analysis of this instability, and more generally 
in studies of the dynamics of astrophysical discs.
\end{abstract}

\section{Introduction}
The magnetorotational instability (MRI; Balbus \& Hawley 1991, 1998; Hawley \& 
Balbus 1991) generates and sustain angular momentum 
transport in differentially rotating astrophysical discs. It does so by 
converting the free energy source contributed by differential rotation into 
turbulent motions (e.g. Balbus 2003), which transport angular momentum via 
Maxwell stresses. 

Most MRI models in non-ideal MHD conditions adopt either the ambipolar 
diffusion (Blaes \& Balbus 1994, MacLow et al. 1995 and Hawley \& 
Stone 1998) or resistive approximations (Jin 1996, Balbus \& Hawley 1998, 
Papaloizou \& Terquem 1997, Sano \& Miyama 1999; Sano, Inutsuka \& Miyama 
1998, Sano et. al. 2000, 
Fleming, Stone \& Hawley 2000 and  Stone \& Fleming 2003). 
The inclusion of Hall 
conductivity terms is a relatively recent development (Wardle 
1999 (W99 hereafter), 
Balbus \& Terquem 2001, Sano \& Stone 2002a,b; 2003, Salmeron \& Wardle 
2003 (SW03 hereafter) and Desch 2004). When the Hall effect dominates over 
ambipolar diffusion, fluid dynamics is dependent on the alignment 
of the magnetic field with the angular velocity vector of the disc (Wardle \& 
Ng 1999), and wave modes supported by the fluid are intrinsically modified. 
For example, left and right-circularly polarised Alfven waves travel at 
different speeds and damp at different rates in this regime (Pilipp et al. 
1987, Wardle \& Ng 1999).
Both the structure and growth rate of MRI perturbations can be substantially 
modified by Hall conductivity, especially when the coupling between ionised 
and neutral components of the fluid is low (W99, SW03).   

In a previous paper we presented a linear analysis of the vertical structure 
and growth of the MRI in weakly ionised, stratified  accretion discs (SW03). 
In that study the components 
of the conductivity tensor were assumed to be constant with height. The 
obtained solutions illustrate the properties of the MRI when different 
conductivity regimes are dominant over the entire cross-section of the disc.
We found that when the magnetic coupling is weak, modes computed with a 
non-zero Hall conductivity grow faster and act over a more 
extended cross-section than those obtained using the ambipolar 
diffusion approximation. The height above the midplane where the 
fastest growing modes peak depends on the conductivity regime of the fluid. 
When ambipolar diffusion is important, perturbations peak 
at a higher $z$ than when the fluid is in the Hall limit.  
Furthermore, when the coupling is weak, perturbations computed with a full 
conductivity 
tensor peak at different heights depending on the orientation of the 
magnetic field with respect to the angular velocity vector of the disc. This 
is a 
consequence of the dependency of the Hall effect on the sign of 
$\bmath{\Omega} \cdot \bmath{B}$ (W99). 
Finally, we showed that in the Hall regime perturbations can have a very 
complex 
structure (high wavenumber), particularly when the magnetic field is weak. In 
these conditions, many modes were found to grow, even with a very weak 
magnetic coupling. These results suggest that 
significant accretion can occur in regions closer to the midplane of 
astrophysical discs, 
despite the low magnetic coupling, due to the large column density of the 
fluid. This idea contrasts with the commonly accepted view that accretion is 
important primarily in the surface regions, where the coupling between 
ionised and neutral components of the fluid is much stronger, but the column 
density is significantly smaller.

In a real disc, the components of the conductivity tensor vary with height 
(e.g. Wardle 2003) as 
a result of changes in the abundances of charged particles, fluid density and 
magnetic field strength. This, in turn, is 
a consequence of changes in the ionisation balance within the disc, which 
reflects 
the equilibrium between ionisation and recombination processes. The former 
are primarily non-thermal, triggered by cosmic rays, x-rays 
emitted by the central protostar and radioactive materials. The later can, in 
general, occur both in the gas phase and in grain surfaces (e.g. Oppenheimer 
\& Dalgarno 1974, Spitzer 1978, Umebayashi \& Nakano 1980, Nishi, Nakano \& 
Umebayashi 1991, Sano et. al. 2000). As a result, different conductivity 
regimes are expected to 
dominate at different heights (Wardle 2003). In this paper we revisit 
the linear growth and structure of MRI perturbations using a height-dependent 
conductivity tensor. We assume that the disc is thin and isothermal and 
adopt a fiducial 
disc model based in the minimum-mass solar nebula (Hayashi 1981, 
Hayashi, 
Nakazawa \& Nakagawa 1985). Further, we assume that ions and electrons are 
the main charge carriers, which is a valid 
approximation in late stages of accretion, after dust grains have settled 
towards the midplane of the disc. As an indication of the timescales for this 
settling to occur, Nakagawa, Nakazawa \& Hayashi (1981) show that the mass 
fraction of 
$\sim 1$ - $10 \micron$ grains well mixed with the gas phase, 
(i.e. not settled), drops from $ \sim 10^{-1} $ to $10^{-4} $ in t 
$ \sim 2 \ee 3 $ to $1 \ee 5 $ years. Furthermore, although the timescale 
for dust 
settling all the way to the equator may exceed the lifetime of the disc, 
grains can settle within a few pressure scaleheights about the 
midplane in a much shorter timescale (Dullemond \& Dominik 2004). 
  
This paper is structured as follows: Section \ref{sec:formulation} summarises 
the formulation, including the governing equations, fiducial disc model and 
ionisation balance. Section \ref{sec:method} describes the linearisation, 
parameters of the problem 
and boundary conditions. Results are presented in sections 
\ref{sec:results1} and \ref{sec:results2}. We summarise in section 
\ref{sec:results1} the test models used in this study and present 
the ionisation rates as a function of $z$ at representative radial locations 
from the central protostar ($R = 1$, $5$ and $10$ AU). The importance of 
different conductivity regimes at different heights is also described for a 
range of magnetic field strengths. Then, in section \ref{sec:results2}, the 
structure and growth rate of MRI perturbations at the 
radii of interest are analysed, including a comparison with results obtained 
using different configurations of the conductivity tensor, sources of
ionisation and disc structure.  We find that the MRI is active 
over a wide range of fluid conditions and magnetic field strengths. For 
example, for the fiducial model at $R = 1$ AU, and including cosmic ray 
ionisation, unstable modes are found for 
$B \lesssim 10$ G. When $200$ mG $\lesssim B \lesssim 5$ G, these 
perturbations grow at 
about the ideal-MHD rate $\sim 0.75$ times the dynamical (keplerian) 
frequency of the disc. Results are discussed in section 
\ref{sec:discussion} and the paper's formulation and key findings are 
summarised in section \ref{sec:summary}.  

\section{Formulation}
\label{sec:formulation}

\subsection{Governing Equations}
\label{subsec:governing}

Following W99 and references therein, we formulate the conservation 
equations 
about a local keplerian frame corotating with the disc at the keplerian 
frequency $\Omega = \sqrt {GM/r^3}$. Time derivatives in this frame, 
 $\partial / \partial t$, correspond to 
$\partial / \partial t + \Omega \partial / \partial \phi$ in the standard 
laboratory system $(r, \phi, z)$ anchored at the central mass M. The fluid 
velocity is expressed as a departure from keplerian motion 
$\bmath{v} = \bmath{V} - \v_K$, where $\bmath{V}$ is the velocity in the 
laboratory frame and $\v_K = \sqrt {GM/r}\  \fh$ is the keplerian
velocity at the radius $r$. The fluid is assumed to be weakly ionised, so 
the effect of ionisation and recombination processes on the neutral gas, as 
well as the ionised 
species' inertia and thermal pressure, are negligible. Under 
this approximation, separate equations of motion for the ionised species are 
not required and their effect on the neutrals is treated via a conductivity 
tensor (W99 and references therein), which is a function of location 
($r$,$z$). 

The governing equations are the continuity equation,

\begin{equation}
\delt{\rho} + \div(\rho \v) = 0 \,,
	\label{eq:continuity}
\end{equation}

\noindent
the equation of motion,

\begin{eqnarray}
\lefteqn{\delt{\v} + (\v \cdot \grad)\v -2\Omega \vf \rh + \half\Omega \vr\fh
-\frac{\vk^2}{r}\rh + \frac{c_s^2}{\rho}\grad\rho + {} }
	\nonumber\\
& & {}+\grad \Phi - \frac{\J\cross\B}{c\rho} = 0\,,
	\label{eq:momentum}
\end{eqnarray}

\noindent
and the induction equation,

\begin{equation}
\delt{\B} = \curl (\v \cross \B) - c \curl \E' -\thalf \Omega \B_r \fh \,.
	\label{eq:induction}
\end{equation}

\noindent
In the equation of motion (\ref{eq:momentum}), $\Phi$ is the gravitational
potential of the central object, given by

\begin{equation}
\Phi = -\frac{GM}{(r^2 + z^2)^{\half}} \,,
          \label{eq:gravpotential}
\end{equation}

\noindent
and $\v_K^2/r$ is the centripetal term generated by exact keplerian
motion. Similarly, $2\Omega \vf\rh$ and $\half \Omega
\vr\fh$ are the coriolis terms associated with the use of a local keplerian
frame, $c_s = \sqrt {P/\rho}$ is the isothermal sound speed,
$\Omega = \v_K/r$ is the keplerian frequency and $c$ is the speed
of light. Other symbols have their usual meanings.

In the induction equation (\ref{eq:induction}), $\E'$ is the electric field 
in the frame comoving with the neutrals and the term $\thalf \Omega \B_r \fh$
accounts for the generation of toroidal field from the radial component due to
the differential rotation of the disc.

Additionally, the magnetic field must satisfy the constraint:

\begin{equation}
\div \B = 0 \,,
\label{eq:divB}
\end{equation}

\noindent
and the current density must satisfy Ampere's law,

\begin{equation}
\J = \frac{c}{4\pi}\grad\cross \B 
\label{eq:j_curlB}
\end{equation}

\noindent
and Ohm's law,

\begin{equation}
	\J = \bmath{\sigma}\cdot \E' = \sigma_{\parallel} \Epa +
	\sigma_1 \Bh \cross \Epe + \sigma_2 \Epe  \,.
	\label{eq:J-E}
\end{equation}

Note that we have introduced the conductivity tensor $\bmath{\sigma}$ in 
equation (\ref{eq:J-E}). We refer the reader to W99 and references therein 
for derivations and additional details of this formulation.
Assuming that the only charged species are ions and
electrons, and that charge neutrality is satisfied
($n_i = n_e$), the components of the conductivity tensor can be expressed 
as (SW03), the conductivity parallel to the magnetic field,

\begin{equation}
	\sigma_\parallel = \frac{cen_e}{B} (\beta_i - \beta_e) \,,
	\label{eq:sigp}
\end{equation}

\noindent
the Hall conductivity,

\begin{equation}
	\sigma_1 = \frac{cen_e}{B} \frac{(\beta_i + \beta_e)(\beta_e -
	\beta_i)}{(1+\beta_e^2)(1+\beta_i^2)} \,,
	\label{eq:sig1}
\end{equation}

\noindent
and the Pedersen conductivity,

\begin{equation}
	\sigma_2 = \frac{cen_e}{B} \frac{(1-\beta_i\beta_e)(\beta_i -
	\beta_e)}{(1+\beta_e^2)(1+\beta_i^2)} \,.
	\label{eq:sig2}
\end{equation}

\noindent
From (\ref{eq:sig1}) and (\ref{eq:sig2}), we find:

\begin{equation}
	\sigma_\perp = \frac{cen_e}{B} \frac{(\beta_i -
	\beta_e)}{[(1+\beta_e^2)(1+\beta_i^2)]^{1/2}} \,.
	\label{eq:sigperp}
\end{equation}

\noindent
where $\sigma_{\perp} = \sqrt{\sigma_1^2 + \sigma_2^2}$ is the total
conductivity perpendicular to the magnetic field. 

In equations (\ref{eq:sigp}) to (\ref{eq:sigperp}),

\begin{equation}
\beta_j= {Z_jeB \over m_j c} \, {1 \over \gamma_j \rho}
\label{eq:Hall_parameter}
\end{equation}
\noindent
is the Hall parameter, given by the ratio of the gyrofrequency and the 
collision frequency of charged species
$j$ with the neutrals. It represents the relative importance of the Lorentz
and drag terms in the charged species' motion. In equation 
(\ref{eq:Hall_parameter}),

\begin{equation}
	\gamma_j = \frac{<\sigma v>_j}{m_j+m} \,,
	\label{eq:gamma}
\end{equation}
\noindent
where $m$ is the mean mass of the neutral particles and $<\sigma v>_j$ is the
rate coefficient of momentum exchange by collisions with the neutrals. 
The ion-neutral momentum rate coefficient is given by,

\begin{equation}
<\sigma v>_i = 1.6\ee -9 \ut cm 3 \ut s -1 \,,
	\label{eq:sigvi} 
\end{equation}

\noindent
an expression that ignores differences in the values of elastic cross-sections 
of 
$\rmn{H}$, $\rmn{H_2}$ and $\rmn{H_e}$ (Draine, Roberge \& Dalgarno 1983). 
Similarly, the rate coefficient of momentum exchange of electrons with the 
neutrals is (Draine et al. 1983): 

\begin{equation}
	<\sigma v>_e \approx 1\ee -15 \left(\frac{128 kT}{9\pi
	m_e}\right)^{1/2} \ut cm -2 \,, \,
	\label{eq:sigve}
\end{equation}

\noindent
and the mean ion mass $m_i = 30 m_H$.

The relative magnitudes of the components of the conductivity tensor 
differentiate three conductivity regimes:
When $\sigma_\parallel \gg \sigma_2 \gg |\sigma_1|$ for most charged species, 
ambipolar diffusion dominates and the magnetic field is effectively frozen 
into the ionised components of the fluid. Electron-ion drift is small compared 
with ion-neutral drift and the ionised species act as a single fluid. MRI 
studies 
in this regime include Blaes \& Balbus (1994), MacLow et al. (1995) and 
Hawley \& Stone (1998). Conversely, when 
$\sigma_\parallel \approx \sigma_2 \gg |\sigma_1|$, the conductivity is a 
scalar, giving rise to the familiar ohmic diffusion, and the magnetic field 
is no longer 
frozen into any fluid component. Examples of studies of the MRI in this 
regime are 
Jin (1996), Balbus \& Hawley (1998), Papaloizou \& Terquem (1997), Sano \& 
Miyama (1999), Sano et. al. (1998, 2000, 2004),
Fleming, Stone \& Hawley (2000) and  Stone \& Fleming, (2003).
Ambipolar diffusion dominates in low density regions, where magnetic stresses
are more important than collisions with the neutrals and the ionised species 
are mainly tied to 
the magnetic field rather than to the neutral particles. On the 
contrary, in relatively high density environments, the ionised species are 
primarily linked to the neutrals via collisions and Ohmic diffusion 
dominates. There is, however, an 
intermediate density range characterised by a varying degree of coupling 
amongst charged species. In these circumstances, there is a component of the 
conductivity tensor that is perpendicular both to the electric and 
magnetic fields. It is this term that gives rise to Hall currents. It has 
been shown that this regime can be important in the weakly ionised 
environment of accretion discs. For 
example, using an MRN grain model (Mathis, Rumpl \& Nordsieck 1977) with a 
power law distribution of grain sizes between $50$ and $2500$ Angstrom, Hall 
conductivity is important for $10^{7} \lesssim n_H \lesssim 10^{11} \percc$
(Wardle \& Ng 1999). 
MRI studies including Hall diffusion have been conducted by W99, Balbus \& 
Terquem (2001), Sano \& Stone (2002a,b; 2003), SW03 and Desch (2004).

The ionisation balance within the disc (section \ref{subsubsec:ionisation}) 
determines the abundances of charged species (ions and electrons). 
These, in turn, determine the configuration of the 
conductivity tensor. In protostellar discs, outside of the central $0.1$ AU, 
ionisation sources are non-thermal (Hayashi 1981) and the ionisation fraction 
is not enough 
to produce good magnetic coupling over the entire cross-section of the disc. 
In these conditions, the region around the midplane is likely to be a 
magnetically ``dead zone'' (Gammie 1996).
 
\subsection{Disc Model}
	\label{subsec:model}

Our model incorporates the vertical structure of the disc, but neglects fluid
gradients in the radial direction. This is an appropriate approximation, as 
astrophysical
accretion discs are generally thin and changes in the radial direction occur
on a much bigger length scale than those in the vertical direction.
Including the vertical structure means that perturbations of spatial
dimensions
comparable to the scale height of the disc, which are associated with either a
strong magnetic field ($v_A \sim c_s$) or low conductivity, can be explored.

We adopted, as our fiducial model, a disc based in the minimum-mass solar 
nebula (Hayashi 1981, Hayashi et. al. 1985).
In this model, the surface density 
$\Sigma(r)$, sound speed $c_s(r)$, midplane density $\rho_o (r)$, scaleheight 
$H(r)$ and temperature $T(r)$ are:

\begin{equation}
\Sigma(r) = \Sigma_o\rau^{-\frac{3}{2}}  \,,
	\label{eq:surf_density}
\end{equation}

\begin{equation}
c_s(r) = c_{so}  \rau^{-\frac{1}{4}}\lau^{\frac{1}{8}} 
\mol^{\frac{1}{2}} \,,
	\label{eq:sound_speed}
\end{equation}

\begin{equation}
\rho_{0}(r) = \rho_o \rau^{-\frac{11}{4}} 
\lau^{-\frac{1}{8}} \mau^{\frac{1}{2}} 
\mol^{\frac{1}{2}} \,, 
	\label{eq:rho_midplane}
\end{equation}

\begin{equation}
H(r) = H_{0} \rau^{1.25} \,, \,
	\label{eq:scaleheight}
\end{equation}

\noindent
and,

\begin{equation}
T(r) = T_o \rau^{-1/2} \lau^{1/4} K \,.
	\label{eq:temperature}
\end{equation} 

\noindent
In the previous expressions,

\begin{equation}
\Sigma_o = 1.7 \ee 3 \, \rmn{g} \, \rmn{cm}^{-2} \,,
	\label{eq:Sigmao}
\end{equation}

\begin{equation}
c_{so} = 9.9 \ee 4 \, \rmn{cm} \, \persec \,,
	\label{eq:cso}
\end{equation}

\begin{equation}
\rho_o = 1.4 \ee -9 \, \rmn{g} \, \percc \,,
	\label{eq:rho_o}
\end{equation}

\begin{equation}
H_0 = 5.0 \ee 11 \u cm \,,
	\label{eq:ho}
\end{equation}

\noindent
and,

\begin{equation}
T_o = 280 \, \rmn{K} \,.
\end{equation}

In equations (\ref{eq:surf_density}) to (\ref{eq:temperature}), $M_\star$ 
and $L_\star$ are the stellar mass and luminosity, respectively, and 
$\mu$ is the mean molecular mass of the gas.  This model describes 
the minimum mass distribution of the solar nebula, estimated assuming an 
efficient planet formation with no significant migration. With these 
assumptions, the current mass 
distribution and composition of the planets is a good indication of that of 
dust in the original nebula. This model has been used extensively, but there 
are theoretical grounds to 
expect that a typical protostellar disc may be more massive, with a 
different large scale structure, shaped ultimately by the action of MHD 
turbulence (e.g. Balbus \& Papaloizou 1999). It can be shown 
that the disc surface density may be 
roughly up to an order of magnitude higher than the one specified by the 
minimum-mass solar nebula 
model before self gravity becomes important. To 
account for the possibility of discs being more massive than the minimum-mass 
solar nebula model, we also studied the properties of MRI unstable 
modes using a disc structure with an increased surface 
density $\Sigma_o^{'} = 10\Sigma_o$ and mass density $\rho_o^{'} = 10\rho_o$.
The Toomre parameter (Toomre 1964) $Q \sim 6$ in this case, so the 
assumption of a non self-gravitating disc is still valid. For simplicity, we 
assume that the temperature $T_o$ is unchanged, so $H_o$ and $c_{so}$ are 
the same as in the fiducial model.    

As the disc is gravitationally stable, the gravitational force in the 
vertical direction comes from the central protostar. Under these 
conditions, the balance between the vertical component of the central
gravitational force and the pressure gradient within the disc determines
its equilibrium structure. The vertical density distribution
in hydrostatic equilibrium is given by

\begin{equation}
\frac{\rho (r,z)}{\rho_{o}(r)}= \exp \left[-\frac{z^2}{2 H^2(r)}\right] \,.
	\label{eq:rhoinitial}
\end{equation}

\noindent
Assuming a neutral gas composed of molecular hydrogen and helium such that 
$n_{He} = 0.2 n_{H_2}$, the neutral gas mass and number densities are, 
respectively,

\begin{equation}
\rho = \Sigma n_i m_i = 1.4 m_H n_H
	\label{eq:comp1}
\end{equation}

\noindent
and

\begin{equation}
n = 1.2 n_{H_2} \, \,,
	\label{eq:n}
\end{equation}

\noindent
which gives,

\begin{equation}
n_{H}(r,z) = \frac{\rho(r,z)}{1.4 m_H}	
	\label{eq:hdens}
\end{equation}

\noindent
and $\mu = \rho/n = 2.34$.
For simplicity, we take $L_\star/L_\odot = M_\star / M_\odot = 2.34 / \mu = 1$
in all our models.

\subsection{Ionisation balance}
	\label{subsubsec:ionisation}

The ionisation balance within the disc is given by the equilibrium between 
ionisation and recombination processes. In general, recombination can take 
place both in the
gas phase and on grain surfaces (e.g. Oppenheimer 
\& Dalgarno 1974, Spitzer 1978, Umebayashi \& Nakano 1980, Nishi, Nakano \& 
Umebayashi 1991, Sano et. al. 2000), but 
here we have assumed that grains have settled, so we are including only 
gas-phase recombination rates.
Except in the innermost sections of the disc ($R \lesssim 0.1$ AU), where 
thermal effects are important (Hayashi 1981), 
ionisation processes in protostellar discs are mainly non-thermal. Ionising 
agents are typically cosmic rays, x-rays emitted by the magnetically active 
protostar, and the decay of radioactive materials present within the disc. 
Some authors (e.g. Fromang, Terquem \& Balbus 2002) have argued that the 
low energy particles important for cosmic-ray ionisation are likely to be 
excluded by the 
protostar's winds. Given the uncertainties involved, and the expectation that 
cosmic rays (if present), may be more important than x-rays near the 
midplane (especially for $R \approx 1$ AU, where the surface density is 
larger than the attenuation length of x-rays), we explore both 
options in this study. The treatment of ionisation and recombination processes
is detailed below.

\subsubsection{Cosmic ray ionisation}
	\label{subsubsec:cosmic}

The cosmic ray ionisation rate, $\zeta_{CR} (r,z)$, is given by,

\begin{eqnarray}
\lefteqn{  \zeta_{CR}(r,z) = \frac{\zeta_{CR}}{2} 
\left[ \rmn{exp}\left(
	-\frac{\Sigma(r,z)}{\lambda_{CR}}\right) +  \right.  }
\nonumber \jot 1.3cm \\
& &  \left.\rmn{exp} \left(- \frac{\Sigma(r) - \Sigma(r,z)}{\lambda_{CR}}
\right)\right] \,,
\label{eq:cosmic_rays}
\end{eqnarray}

\noindent
where $\zeta_{CR} = 10^{-17} \persec$ is the ionisation rate due to 
cosmic rays in the 
interstellar medium and $\lambda_{CR} = 96$ gr $\, \rmn{cm}^{-2}$ is the 
attenuation length, a measure of how deep cosmic rays can penetrate the disc 
(Umebayashi \& Nakano 1981). When the surface density is larger than $\sim 2 
\lambda_{CR}$, most cosmic rays do not reach the midplane. Also,

\begin{equation}
\Sigma(r,z) = \int_{z}^{\infty} \rho(r,z) dz \,,
	\label{eq:attenuation}
\end{equation} 

\noindent
is the vertical column density from the location of interest to 
infinity. The two terms in brackets in equation (\ref{eq:cosmic_rays}) measure 
the ionisation rate at the position ($r$, $z$) by cosmic rays penetrating the 
disc from above, and below, respectively.

\subsubsection{X-ray ionisation}
	\label{subsubsec:xray}

There is strong evidence for an enhanced magnetic activity 
in young stellar objects (i.e. see review by Glassgold, Feigelson \& Montmerle 
2000). Typical soft x-ray luminosities  ($0.2$ - $2$ kev) of these objects are 
in the range of $10^{28} $ - $10^{30} \rmn{erg} \persec$, about $10^{2} $ - 
$10^{3}$ times more energetic than solar levels (Glassgold et. al. 2000). 
Models of the penetration of stellar x-rays into a protostellar disc by 
Igea \& Glassgold (1999), show that even discounting the low energy photons 
that are 
attenuated by stellar winds, the ionisation rate due to hard x-rays close to 
the central object is many orders of magnitude larger than 
that of cosmic rays, especially above $z/H \sim 2$. These authors 
investigated the 
ionisation rate by x-rays from a central protostar modelled as a x-ray source 
of total luminosity $L_x \sim 10^{29}\rmn{erg} \persec$ and temperature $kT_x$ 
in the range of $3$ - $8$ kev. The transport of x-rays through the disc was 
followed using a Monte Carlo procedure which included both absorption and 
scattering by disc material. The incorporation of 
scattering is important, as x-rays can be scattered not only out of the 
disc, 
but also towards the midplane, enhancing the ionisation level deeper within 
the disc. Results indicate that, at each radius, the x-ray 
ionisation rate is a function of the vertical column density into the disc 
$N_{\perp}(r,z)$, irrespective of its structural details.

To calculate the x-ray ionisation rate $\zeta_{X}(r,z)$ in the upper half of 
the disc we added the contribution of x-rays arriving from both sides. 
We used the values of $\zeta_{X}(r,z)$ ($\persec$) as a function of the 
vertical column density, $N_{\perp} \, (\rmn{cm}^{-2})$, plotted in Fig. 3 of 
Igea \& Glassgold (1999), for $R = 1$, $5$ and 
$10$ AU and $kT_x = 5$ Kev. The vertical column density, appropriate for 
x-rays arriving from the top is 
 
\begin{equation}
N_{\perp}(r,z) = \int_{z}^{\infty} n_H(r,z) dz \,, \,
	\label{eq:nperp}
\end{equation}

\noindent
with $n_H$ given by equation (\ref{eq:hdens}). Similarly, the 
ionisation 
contributed by x-rays arriving at ($r,z>0$) from the other hemisphere of the 
disc is obtained substituting $N_{\perp}$ above by $2N_{\perp}(r,0) - 
N_{\perp}(r,z)$, although this contribution is usually negligible.

\subsubsection{Radioactivity}
	\label{subsubsec:radioactivity}

The ionisation rate $\zeta_{R}$ contributed by the decay of radioactive 
materials (primarily $\,^{40}K$) have also been considered 
in previous work (i.e. Consolmagno \& Jokipii 1978, Sano et al. 2000). This 
rate can be calculated as (Consolmagno \& Jokipii 1974): 

\begin{equation}
	\label{eq:radio}
\zeta_R = \frac{\gamma n_r E}{\Delta \varepsilon} \,,
\end{equation}

\noindent
where $\gamma$ \ ($\persec$) is the decay rate of the radioactive isotopes, 
$n_r$ their number density relative to 
hydrogen, and $E$ (eV) the energy of the produced radiation. Similarly, 
$\Delta \varepsilon = 37$ eV, is the average energy for the production of an 
ion pair in $H_2$ gas (Consolmagno \& Jokippi 1974 and references 
therein, Shull \& Van Steenburg 1985, Voit 1991).  In 
the present study we have adopted,

\begin{equation} 
\zeta_R = 6.9 \ee -23 [\delta_2 + (1 - \delta_2) f_g ] \, \persec 
	\label{eq:radrate}
\end{equation}

\noindent
(Umebayashi \& Nakano 1981, 1990). Here, 
$\delta_2 \approx 0.02$ is the fraction of heavy metal atoms in the gas 
phase, 
estimated via measurements of interstellar absorption lines in diffuse clouds 
(Morton 1974). On the other hand, $f_g$ is a parameter that takes into 
account 
the degree of sedimentation of dust grains in protostellar discs with 
respect to interstellar values (Hayashi 1981, Sano et. al. 2000). 

Although the ionisation effect 
contributed by this agent is very small compared with that of x-rays and 
cosmic rays, we included it because it may well be the only mechanism active 
in regions close to the midplane (particularly for $R \lesssim 5$ AU) in the 
scenario where 
cosmic rays are excluded from the disc. In this case, it is 
important to explore the sensitivity of MRI perturbations to changes in the 
level of depletion of dust grains ($f_g$). 

We have 
assumed here that the ionised component of the fluid is made of 
ions and electrons only, a case that corresponds to late evolutionary stages 
of protostellar discs, after dust grains have settled towards the midplane. 
Accordingly, we assume $f_g = 0$ in all our models, except when 
especifically exploring the role of radioactivity in a disc where cosmic rays 
are assumed to be excluded from it.

\subsubsection{Recombination rate}
	\label{subsubsec:recomb}

Gas-phase recombination occurs through dissociative recombination  of 
electrons with molecular ions and, at a slower rate, via radiative 
recombination with metal ions. It has been pointed out by previous authors 
that the ionisation balance may be especially sensitive to the presence of 
metal atoms within the disc. For some disc configurations, a number density 
of metals as small as $10^{-7} $ times the cosmic abundance may be enough to 
make 
turbulent the whole cross-section of the disc, eliminating the central 
magnetic dead zone (Fromang et. al. 2002). This is so because metal atoms 
generally take the charges of molecular ions, but recombine with 
electrons (via radiative processes) at a much more slower rate.

If $n_m^{+}$ and $n_M^{+}$ are the number densities of molecular and metal 
ions, respectively, the rate equations for $n_e$ and $n_m^{+}$ can be 
expressed as (e.g. Fromang et. al. 2002):

\begin{equation}
\frac{d n_e}{dt} = \zeta n_H - \beta n_e n_m^{+} - \beta_r n_e n_M^{+}
	\label{eq:nerate}
\end{equation} 
 
\noindent
and

\begin{equation}
\frac{d n_m^{+}}{dt} = \zeta n_H - \beta n_e n_m^{+} - \beta_t n_M n_m^{+} \,,
	\label{eq:nmrate}
\end{equation}

\noindent
where $\zeta$ is the total 
ionisation rate, calculated as summarised in the previous sections, and 
$n_\mathrm{H}$ is the hydrogen number density from equation (\ref{eq:hdens}).

In the previous expressions, $\beta$ is the dissociative recombination rate 
coefficient for molecular ions, $\beta_r$ is the radiative recombination rate 
coefficient for metal ions and $\beta_t$ is the rate coefficient of charge 
transfer from 
molecular ions to metal atoms. If all metals are locked into dust grains, 
which have in turn sedimented towards the midplane of the disc, equations 
(\ref{eq:nerate}) and (\ref{eq:nmrate}), together with charge neutrality, 
$n_e = n_M^{+} + n_m^{+}$, and appropriate values for the rate coefficients,

\begin{equation}
\beta_r = 3 \ee -11 T^{-1/2} \, \rmn{cm}^{3} \persec \,,
	\label{eq:betar}
\end{equation} 

\begin{equation}
\beta = 3 \ee -6 T^{-1/2} \, \rmn{cm}^{3} \persec \,,
	\label{eq:beta}
\end{equation}

\begin{equation}
\beta_t = 3 \ee -9 \rmn{cm}^{3} \persec \,,
\end{equation}

\noindent
(see Fromang et al 2002 and references therein), lead to

\begin{equation}
n_e \approx \sqrt{\frac{\zeta n_H}{\beta}} \,.
	\label{eq:ne}
\end{equation}

\noindent
However, as these authors point out also, dust grains not only absorb metal 
atoms, but also release them as a result of the action of x-rays. Because of 
this effect, the 
abundance of metal atoms in a disc could be quite insensitive to the spatial 
distribution of dust grains. This, in turn, means that dust settling does 
not necessarily lead to a severe reduction in 
the number density of metal atoms in the gas phase. If these dominate, the 
corresponding $n_e$ is

\begin{equation}
n_e \approx \sqrt{\frac{\zeta n_H}{\beta_r}} \,.
	\label{eq:ne2}
\end{equation}

\noindent
The transition from one regime to the other may be shown to occur when
(Fromang et. al. 2002), 

\begin{equation}
x_M \approx 10^{-2} T^{-1/2} x_e \,. 
      \label{eq:transition}
\end{equation}

It is clear then, that the evolutionary 
stage --and activity-- of the disc are important factors in the ionisation 
balance, 
as they influence the degree of sedimentation of dust grains. As the disc 
evolves, dust grains tend to occupy a thin layer around the midplane, 
becoming removed --in any dynamical sense-- from the gas at higher vertical
locations. This causes the ionisation fraction of the gas to increase, by 
eliminating recombination pathways on dust surfaces. In the present work, we 
have used equation (\ref{eq:ne2}) to calculate the electron (and ion) number 
density. The 
minimum values of $x_M$ for this approximation to be valid 
(equation (\ref{eq:transition})) have also been 
computed and compared with an estimate of metal abundances in the gas phase 
(see section \ref{subsec:ionrate}). Results indicate 
that, for the range of parameters adopted here, the abundance of metal atoms
is such that radiative recombination is indeed dominant, so the use of 
equation  (\ref{eq:ne2}) to calculate the electron fraction is justified. 

\section{METHODOLOGY}
\label{sec:method}

\subsection{Linearisation}
	\label{subsec:linear}

Full details of the methodology are described in SW03. For the sake of 
clarity, and completeness, we summarise here 
the most important steps and point at some differences with the previous
formulation. The system of 
equations (\ref{eq:continuity}) to (\ref{eq:induction}), (\ref{eq:j_curlB}) 
and
(\ref{eq:J-E}) was linearised about an initial steady state where fluid 
motion is exactly keplerian and the magnetic field is vertical,
so $\J = \v = \E' = 0$ and $\B = B \hat{z}$. 
In the initial state both $\E'$ and $\J$ vanish, so the changes in the
conductivity tensor due
to the perturbations do not appear in the linearised equations. As a result,
it is not necessary to explore how the perturbations affect the
conductivity of the fluid and only the values in the initial steady state 
are required.

We assume that the wavevector of the perturbations is perpendicular to the 
plane
of the disc ($k = k_z$). These perturbations, initiated from a vertical 
magnetic field, are the fastest growing modes when the fluid is in either the 
Hall or resistive conductivity regimes, as in these cases magnetic pressure 
suppresses 
displacements with a non-zero radial wavenumber (Balbus \& Hawley 1991, Sano 
\& Miyama 1999). However, as pointed out by Kunz \& Balbus (2004)
through their analysis of local axisymmetric MRI, 
this is not necessarily the case for perturbations in the ambipolar diffusion
limit. Their formulation includes radial and vertical wavenumbers as well as 
vertical and azimuthal field components. Results indicate that in this 
regime, 
the fastest growing modes may exhibit both radial and vertical wavenumbers. 
Interestingly, in this configuration unstable modes are present even for 
outwardly increasing angular velocity profiles, contrary to the case in 
ideal-MHD. 
 
Taking perturbations of the form
$\bmath{q} = \bmath{q}_{0} + \bmath{\delta q}(z) \e^{i\omega t}$ about the
initial state, linearising
and neglecting terms of order $H/r$ or smaller, we find that the final linear 
system of equations that describes the MHD perturbations within the disc is,

\begin{equation}
i\omega\rho \delta v_r - 2 \rho \Omega \delta v_{\phi} - 
\frac{B_0}{c}\delta J_{\phi} = 0 \,,
	\label{eq:motion_r_final}
\end{equation}

\begin{equation}
i\omega\rho \delta v_{\phi} + \half\rho \Omega \delta v_r + 
\frac{B_0}{c} \delta J_r = 0 \,,
	\label{eq:motion_phi_final}
\end{equation}

\begin{equation}
i\omega \delta B_r - c\frac{d \delta E_{\phi}}{dz} = 0 \,,
	\label{eq:induc_r_final}
\end{equation}

\begin{equation}
i\omega \delta B_{\phi} + c\frac{d \delta E_r}{dz} + 
\thalf\Omega \delta B_r = 0 \,,
	\label{eq:induc_phi_final}
\end{equation}

\begin{equation}
\delta J_r = - \frac{c}{4 \pi} \frac{d \delta B_{\phi}}{dz} \,,
	\label{ampere_r_final}
\end{equation}

\begin{equation}
\delta J_{\phi} = \frac{c}{4\pi} \frac{d \delta B_r}{dz} \,,
	\label{eq:ampere_phi_final}
\end{equation}

\vspace{0.14cm}

\begin{equation}
\delta J_r = \sigma_2 \delta E_r' - \sigma_1\delta E_{\phi}' \,,
	\label{eq:ohm_r_final}
\end{equation}

\vspace{0.14cm}

\begin{equation}
\delta J_{\phi} = \sigma_1 \delta E_r' + \sigma_2\delta E_{\phi}' \,,
	\label{eq:phm_phi_final}
\end{equation}

\noindent
where $\delta E_{\phi}$ and $\delta E_r$ are the perturbations of the electric
field in the laboratory frame,

\begin{equation}
\delta E_{\phi} = \delta E_{\phi}' + \frac{B_0}{c}\delta v_r \,, \textrm{\ and}
	\label{eq:E'phi_v}
\end{equation}

\begin{equation}
\delta E_r = \delta E'_r - \frac{B_0}{c}\delta v_{\phi} \,.
	\label{eq:E'r_v}
\end{equation}

Note that $\sigma_{\parallel}$, the component of the 
conductivity tensor parallel to the magnetic field, does not appear in the 
linearised equations.
This, in turn, implies that in the linear phase of the MRI and under the 
adopted approximations, the ambipolar diffusion and resistive conductivity 
regimes behave identically (see section \ref{subsec:governing}).

We express equations (\ref{eq:motion_r_final}) to (\ref{eq:E'r_v}) in
dimensionless form by normalising the variables as follows:

\begin{displaymath}
z^{*} = \frac{z}{H} \qquad
\rho^{*} = \frac{\rho(r,z)}{\rho_0(r)} \qquad
\dB^* = \frac{\dB}{B_0} \qquad
\end{displaymath}

\vspace{0.3cm}

\begin{displaymath}
\dv^* = \frac{\dv}{c_s} \qquad
\dE^* = \frac{c \dE}{c_s B_0} \qquad
\dE'^* = \frac{c \dE'}{c_s B_0} \qquad
\end{displaymath}

\vspace{0.3cm}

\begin{displaymath}
\dJ^* = \frac{c \dJ}{c_s B_0\sigma_{\perp}} \qquad
\bmath{\sigma}^{*} = \frac{\bmath{\sigma}}{\sigma_{\perp}} \qquad
\end{displaymath}

Here, subscript `o' denotes variables at the midplane of the disc. Note that 
we have used the local $\sigma_{\perp}$ instead of $\sigma_{\perp_0}$, as in 
SW03, to normalise $\bmath{\sigma}$ and $\dJ$. This is more useful when 
dealing with a height-dependent conductivity. In the following 
dimensionless system, we have dropped the asterisks and expressed the final 
equations in matrix form:

\begin{equation}
\frac{d}{dz}\left( \begin{array}{c}
B_r \\
\\
B_{\phi} \\
\\
E^{}_r \\
\\
E^{}_{\phi}
\end{array} \right)
 = \left( \begin{array}{cccc}
0 & 0 & C_1 A_1 & C_1 A_2 \\
\\
0 & 0 & -C_1 A_2 & C_1 A_3 \\
\\
-\thalf & -\nu & 0 & 0\\
\\
\nu & 0 & 0 & 0
\end{array} \right) \left( \begin{array}{c}
B_r\\
\\
B_{\phi}\\
\\
E_r\\
\\
E_{\phi}
\end{array} \right)
	\label{eq:matrix}
\end{equation}

\vspace{10pt}

\begin{equation}
\dJ = C_2 \left( \begin{array}{cc}
A2 & -A3\\
\\
A1 & A2
\end{array} \right) \dE
	\label{eq:matrix_JE}
\end{equation}

\vspace{10pt}

\begin{equation}
\dv = \chi \frac{1}{1 + \nu^2}
\left( \begin{array}{cc}
-2 & \nu\\
\\
-\nu & -\half
\end{array} \right) \dJ
	\label{eq:matrix_vJ}
\end{equation}

\vspace{10pt}

\begin{equation}
\dE' = \frac{1}{\sigma_{\perp}}\left( \begin{array}{cc}
\sigma_2^{} & \sigma_1^{}\\
\\
-\sigma_1^{} & \sigma_2^{}
\end{array} \right)\dJ
	\label{eq:matrix_EJ}
\end{equation}

\noindent
where

\vspace{8pt}

\begin{equation}
\nu =\frac{i\omega}{\Omega} \,,
	\label{eq:nu}
\end{equation}

\begin{equation}
C_1 = \chi \left(\frac{v_A}{c_s}\right)^{-2} C_2 \,,
	\label{eq:C1}
\end{equation}

\begin{equation}
C_2 = \left[1+\chi\frac{1}{1+\nu^2}
\left(\frac{5}{2}\frac{\sigma_1}{\sigma_{\perp}} +
2 \nu \frac{\sigma_2}{\sigma_{\perp}} + \chi
\right)\right]^{-1} \,,
	\label{eq:C0}
\end{equation}

\begin{equation}
A_1 = \frac{\sigma_1}{\sigma_{\perp}} + 2\chi
\frac{1}{1 + \nu^2} \,,
	\label{eq:A1}
\end{equation}

\begin{equation}
A_2 = \frac{\sigma_2}{\sigma_{\perp}} + \nu \chi
\frac{1}{1 + \nu^2} \,, \textrm{\ and}
	\label{eq:A2}
\end{equation}

\begin{equation}
A_3 = \frac{\sigma_1}{\sigma_{\perp}} + \frac{1}{2}\chi
\frac{1}{1 + \nu^2} \,.
	\label{eq:A3}
\end{equation}

\noindent
In the above expressions,

\begin{equation}
v_A = \frac{B_o}{\sqrt {4 \pi \rho}}
	\label{eq:alfven}
\end{equation}

\noindent
is the local Alfv\'en speed in the disc, and

\begin{equation}
\chi \equiv \frac{\omega_{c}}{\Omega} \equiv \frac{1}{\Omega}\frac{B_o^2
\sigma_{\perp}}{\rho c^2}
	\label{eq:chi}
\end{equation}

\noindent
controls the local coupling between the magnetic
field and the disc (W99, SW03, see also section \ref{subsec:param}). 
Non-ideal effects strongly modify wavemodes at, or above, the critical 
frequency $\omega_c$.

\subsection{Parameters}
	\label{subsec:param}

The following parameters control the dynamics and evolution of the fluid:

\begin{enumerate}
\item $v_A/c_s$, the ratio of the local Alfv\'en speed to the isothermal sound
speed of the gas. It is a measure of the strength of the
magnetic field. In ideal-MHD conditions, unstable modes grow when the 
magnetic field is subthermal ($v_A/c_s < 1$). Under this approximation, when 
$v_A \sim c_s$ the minimum
wavelength of the instability is of the order of the scale height of the disc
and the growth rate decreases rapidly. This is also the case under the 
assumption that ambipolar or ohmic diffusion dominates over the entire 
cross-section of the disc (SW03). However, when the fluid is in the Hall 
regime, with the magnetic field counteraligned with the angular velocity 
vector of the disc ($\bmath{\Omega} \cdot \bmath{B} < 0$), MRI unstable 
modes may exist for stronger fields, up to $v_A/c_s \lesssim 3$ (SW03). 

\item $\chi$, a parameter that characterises the strength of the local 
coupling
between the magnetic field and the disc (equation
(\ref{eq:chi})). It is given by the ratio of the critical frequency above
which flux-freezing
conditions break down and the dynamical (keplerian) frequency of the disc. If 
$\chi \equiv \omega_{c}/\Omega < 1$, the
disc is poorly coupled to the disc at the
frequencies of interest for dynamical analysis, which are also the 
interesting frequencies for the study of the MRI.

\item $\sigma_1/\sigma_2$, the ratio of the conductivity terms
perpendicular to the magnetic field. It is an indication of the
conductivity regime of the fluid, as discussed in section
\ref{subsec:governing} .
\end{enumerate}

We calculated the values of the three parameters above at different locations 
($r$, $z$) within the disc using equations (\ref{eq:sig1}), (\ref{eq:sig2}), 
(\ref{eq:sound_speed}), (\ref{eq:alfven}) and (\ref{eq:chi}). The strength of 
the magnetic field is a free parameter of the problem.

\subsection{Boundary Conditions}
	\label{subsec:boundary}

To solve equations (\ref{eq:matrix}) to (\ref{eq:matrix_EJ}) it is necessary
first to integrate the system of ordinary differential equations (ODE) in
(\ref{eq:matrix}).
This can be treated as a two-point boundary value problem for coupled
ODE. Five boundary conditions must be formulated, prescribed either at the
midplane or at the surface of the disc.

\begin{description}
\item \textbf{At the midplane}: We chose to assign odd symmetry to
$\delta B_r$ and $\delta B_{\phi}$, so they vanish at $z=0$. This gives us
two boundary conditions at the midplane. Also, as the equations are linear, 
their overall scaling is arbitrary, and a
third boundary condition can be obtained by setting one of the
fluid variables to any convenient value. To that effect, we assigned
a value of $1$
to $\delta E'_r$. The three boundary conditions applied at the midplane are, 
then:

\begin{displaymath}
\dB_r = \dB_{\phi} = 0, \textrm{\ and}
\end{displaymath}

\begin{displaymath}
\dE_r' = 1 \,.
\end{displaymath}

\item \textbf{At the surface}: $\chi$ is inversely proportional to the 
density (equation \ref{eq:chi}), so if the conductivity is assumed to be 
constant with height, it 
increases monotonically with $z$. This was the case 
in SW03, where we used this argument to propose that at sufficiently high $z$ 
above the midplane, ideal MHD conditions held. This, in turn, implies that 
the wavelengths of magnetic field perturbations, given the adopted dependency 
of $\rho$ 
with $z$ (equation (\ref{eq:rhoinitial})), must tend to infinity when 
$z \rightarrow \infty$. As a result, the amplitude of such modes should 
vanish at infinity as well. Here, however, the conductivities vary with $z$, 
so 
this argument needs revisiting. In section \ref{subsubsec:regimes} we analyse 
the dependency of $\chi$ with height for different radial positions and a 
range of magnetic field strengths. Here, we just highlight that at the 
surface of the disc ($z/H = 6$, see below), 
the magnetic coupling is still strong. For example, for the fiducial model at
$R = 1$ AU, $\chi \sim 20$. It decreases to $\sim 4.5$ for $R = 5$ AU and to 
$\sim 2.5$ for $R = 10$ AU. This is still above the limit ($\sim 1$) for 
strong coupling (W99). 

When $\chi>10$ the growth rate and characteristic wavenumber of local
unstable modes differ little from the ideal limit (W99), so the same line of 
reasoning of SW03 can be used to argue that in this case $\delta B_r$ and 
$\delta B_{\phi}$ should be zero at the boundary as well. Although $\chi$ at 
the surface when $R \gtrsim 5$ AU is below $10$, it is considered that the 
magnetic coupling there is still sufficiently strong for the ideal-MHD 
approximation to be essentially valid as well. This is also confirmed  by the 
way in which magnetic field perturbations tend to zero at the boundary at 
these radii (see results in sections \ref{sec:results1} and 
\ref{sec:results2}). Consistently with this, both
$\dE'$ and $\dJ$ vanish at the surface as well.
This gives us the remaining two boundary conditions required to integrate
the system of equations (\ref{eq:matrix}). 

We chose to locate the boundary at $z/H = 6$, after confirming that 
increasing this height does 
not significantly affect either the structure or the growth rate of unstable 
modes.  Summarising, the boundary conditions adopted at the surface are:

\begin{displaymath}
\dB_r = \dB_{\phi} = 0 
\end{displaymath}

\noindent
at $z/H = 6$.
\end{description}

This system of equations is solved by `shooting' from the midplane to the 
surface of the disc while simultaneously adjusting the growth rate $\nu$ and 
$\dE'_{\phi}$. The adjustment is done via a 
multidimensional, globally convergent Newton-Raphson method, until the 
solution converges.

\section{DISC CONDUCTIVITY}
\label{sec:results1}

\subsection{Test Models}
\label{subsec:test}

Our fiducial model is based in the 
minimum-mass solar nebula disc (Hayashi 1981, Nakazawa \& Nakagawa 1985; see 
section \ref{subsec:model}). 
The structure and growth rate of MRI unstable 
modes are calculated at representative radial positions ($R = 1$, $5$ and $10$ 
AU) from the central protostar. Two scenarios are explored: Cosmic rays 
either penetrate the disc, although attenuated as 
appropriate (as given by eq. (\ref{eq:cosmic_rays})), or they are 
excluded from it 
by the winds produced by the central object. Unless stated otherwise, results 
presented using this model include cosmic ray ionisation. 
We also consider a disc model with an increased mass and surface density, as 
described in
section \ref{subsec:model}. For simplicity, other disc parameters 
remain unchanged  and in this case, all calculations 
incorporate the ionisation rate provided by cosmic rays. 

We also compare solutions obtained using different configurations of the 
conductivity tensor. By comparing the structure and growth rate of unstable 
modes found using the commonly adopted ambipolar diffusion --or resistive-- 
($\sigma_1 = 0$) 
approximations with those obtained with a full conductivity tensor 
($\sigma_1B_z > 0$, $\sigma_2 \not= 0$), we can appreciate how, and in 
which 
regions of parameter space, Hall conductivity alters the properties of the 
instability. This is explored
for the fiducial model at $1$ AU only.

The properties of the MRI in the Hall limit are 
dependent on the alignment of the magnetic field and angular velocity 
vectors of the disc (W99). The case when these vectors are parallel 
(antiparallel) is characterised by $\sigma_1B_z > 0$ ($\sigma_1B_z < 0$). 
In the Hall ($\sigma_1B_z < 0$) limit, our code 
fails to converge whenever the combination of parameters is such that 
$\chi < 2$ anywhere in the domain of integration. This 
is not surprising as, in this regime, when 
$0.5 < \chi < 2$ all wavenumbers are unstable (W99). 
As a result, we explored the effect of the alignment of $\B$ and 
$\bmath{\Omega}$ by comparing 
solutions obtained with a full conductivity tensor but incorporating 
$\sigma_1B_z$ terms of opposite sign. Even with this approach, solutions 
including a 
$\sigma_1B_z < 0$ term could not be computed at all for $R = 1$ 
AU, given that the midplane coupling at this radius is extremely low 
($\sim 10^{-10}$ to $10^{-2}$ for $B$ between $1$ mG and $1$ G). Therefore, we
present
this analysis at $5$ and $10$ AU only. Solutions incorporating a 
negative Hall conductivity could be obtained for $B \gtrsim 8$ mG at $5$ AU 
and $B \gtrsim 1.5$ mG at 10 AU, where 
the magnetic coupling at the midplane is $\gtrsim 2$. 

Unless stated otherwise, results discussed in 
the following sections incorporate a $\sigma_1B_z > 0$ Hall term contribution.

\subsection{Ionisation Rates}
\label{subsec:ionrate}

\begin{figure} 
\centerline{\epsfxsize=8cm \epsfbox{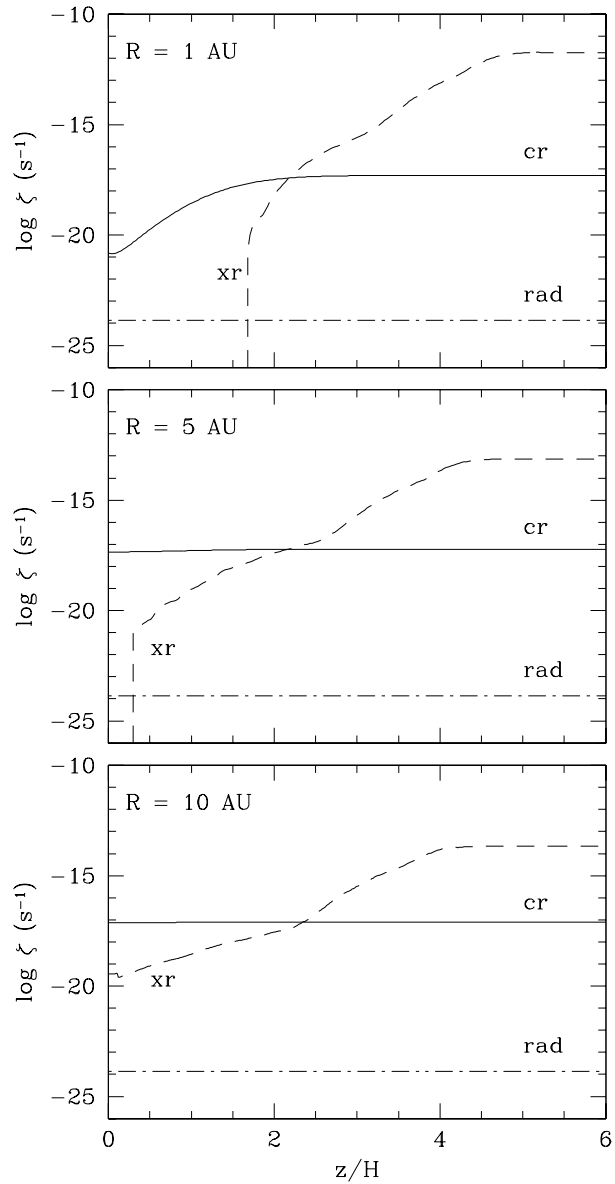}}\vskip 0cm
\caption{Ionisation rate ($\persec$) contributed by cosmic rays (curve 
labeled `cr'), 
x-rays (xr) and radioactive materials (rad) as a function of height, for the 
fiducial model (minimum mass solar nebula disc). Results are shown for 
$R = 1$, $5$ and $10$ AU. Note that at $1$ 
AU, cosmic ray ionisation is attenuated with respect to the interstellar rate 
for $z/H \lesssim 2$. Also, x-rays are excluded from the 
disc for $z/H \lesssim 1.7$ (at $1$ AU) and $z/H \lesssim 0.3$ (at $5$ AU). 
For $R = 10$ AU, 
they reach the midplane, although their ionisation rate is heavily 
attenuated. Cosmic rays 
(if present) are the most efficient ionising agent for $z/H \lesssim 2.2$ in 
all cases.}
\label{fig:ionbalance}
\end{figure}

Fig. \ref{fig:ionbalance} shows the ionisation rate contributed by cosmic 
rays, x-rays and radioactive materials as a function of height for the 
fiducial model. Results are presented at the radial locations of interest: 
$1$, $5$ and $10$ AU. Note that x-rays penetrate up to 
$z/H \sim 1.7$ and $z/H \sim 0.3$ for $R = 1$ and $5$ AU, respectively. 
They reach the 
midplane for $R = 10$ AU. The x-ray ionisation rate is heavily 
attenuated below $\sim 1.5$ to $2$ scaleheights from the surface, depending 
on the radius.
 
At $1$ and $5$ AU, the regions where x-rays are not able to penetrate are 
ionised only by the action of cosmic rays 
(if present) and radioactive decay. Moreover, cosmic rays constitute the most 
efficient ionising agent for $z/H \lesssim 2.2$ at all studied radii. As a 
result, if they are excluded from the disc by the protostar's winds 
(i.e. Fromang et al. 2002), the magnetic 
coupling of the gas in these sections of the disc is expected to be severely 
reduced. In
order to explore the properties of MRI perturbations under this assumption, 
we compare results  computed including and excluding this source of 
ionisation for the fiducial model (section \ref{sec:results2}). 
 
Igea \& Glassgold (1999) have pointed out that in protostellar discs, the 
x-ray ionisation rate can be $10^3$ - $10^5$ 
times that of interstellar cosmic rays. Their calculations show 
that x-rays 
dominate over cosmic rays until the vertical column density, $N_{\perp}$, is 
$10^{25}$, 
$5 \ee 23 $ and $2 \ee 23 \persqcm$ for $R = 1$, $5$ and $10$ AU, 
respectively. Once $N_{\perp}$ exceeds $\sim 10^{25} \persqcm$, x-ray 
ionisation dies out because there are no remaining incident photons available.
The ionisation rates adopted in this study (Fig. \ref{fig:ionbalance}) are in 
agreement with these findings.  

\begin{figure} 
\centerline{\epsfxsize=8cm \epsfbox{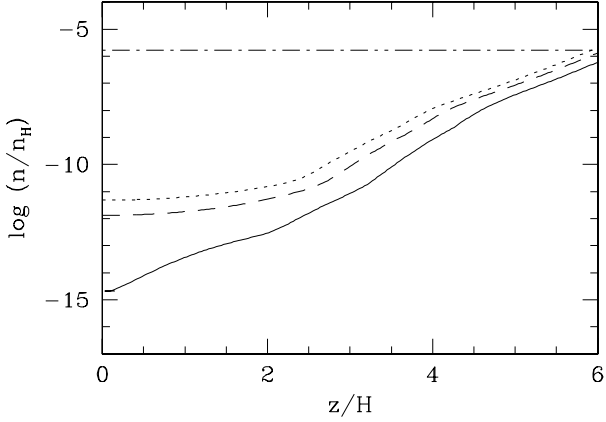}}\vskip 0cm
\caption{Comparison of the minimum abundance of metal atoms in the 
gas phase, $x_M$ ($n/n_H$), 
for the radiative recombination rate of metal ions ($\beta_r$) to be the 
dominant recombination mechanism (Fromang et. al. 2002), and an estimate of 
the total gas-phase metal abundance, 
as a function of height. Minimum abundances 
are shown for the fiducial model at $1$ AU (solid line), $5$ AU (dashed line) 
and $10$ AU (dotted line).
The estimated total gas-phase metal abundance (dot-dashed line) is from 
Umebayashi \& Nakano (1990). 
This abundance exceeds the minimum required for all 
radii and vertical locations of interest in this study.}
\label{fig:ionbalance1}
\end{figure}

We have used eq. (\ref{eq:ne2}) to calculate the electron number density 
(section \ref{subsubsec:recomb}). This expression is valid when metal atoms 
dominate and electrons recombine primarily via radiative processes. Here, 
we justify this choice. Fig. \ref{fig:ionbalance1} compares 
the minimum abundance of metals, (eq. (\ref{eq:transition})), for this 
approximation to be valid with an estimate of total metal abundances in the 
gas phase.
Minimum abundances (relative to hydrogen)  are shown for the radial locations
of interest for the fiducial model. The total gas-phase metal 
abundance is taken as $8.4 \ee -5 \delta_2$ (Umebayashi \& 
Nakano 1990). From Fig. \ref{fig:ionbalance1} it is clear that the abundance 
of gas-phase metal atoms is expected to exceed the minimum for
radiative recombination processes to be dominant at all radii of interest 
here. As a result, we conclude that equation 
(\ref{eq:ne2}) is a valid approximation to calculate the electron number 
density in this study.

\subsection{Magnetic coupling and conductivity regimes as a function of 
height}
\label{subsubsec:regimes}

It is interesting to explore which conductivity terms (section 
\ref{subsec:governing}) are dominant at different heights at the 
representative radial locations we are considering. This information will be 
used in the analysis of the structure and growth of MRI unstable modes 
in the next section. We 
recall that Hall diffusion is locally dominant when (SW03),

\begin{equation}
\chi < \chi_{crit} \equiv \frac{|\sigma_1|}{\sigma_{\perp}} \,.
	\label{eq:criterion}
\end{equation}

\noindent
When $\chi$ is higher than this, but still $\lesssim 10$, ambipolar diffusion 
dominates. In regions where $\chi$ is stronger, ideal-MHD conditions hold 
(W99). We remind the reader than in the present formulation, the ambipolar 
and Ohmic conductivity 
regimes behave identically. This is the case because $\sigma_{\parallel}$ 
(the conductivity 
component that distinguishes between them), does not appear in the final, 
linearised 
system of equations (see section \ref{subsec:linear}). This should be kept in 
mind when analysing the results, as even though the case where $\sigma_1 = 0$ 
has been referred to as the `ambipolar diffusion' limit  
($\sigma_{\parallel} \gg \sigma_2$), it is also consistent with the 
`resistive' regime ($\sigma_{\parallel} \sim \sigma_2$). 

We are interested in 
comparing $\chi$ and $\chi_{crit}$ as a function 
of height for different disc models, radial locations and choices of the 
magnetic field strength. An 
example of the typical dependency of these variables with $z$ is shown in 
Fig. \ref{fig:ionexample} for the fiducial model at 
$R = 1$ AU for $B = 10$ mG. Note that the 
conductivity term parallel to the field increases monotonically with $z$ and 
is independent of $B$ (see eqs. (\ref{eq:sigp}) and 
(\ref{eq:Hall_parameter})). In this case, $\sigma_1$ is nearly two 
orders 
of magnitude smaller than $\sigma_2$ at the midplane. As a result, 
near the midplane $\sigma_{\parallel} \sim \sigma_2 \gg \sigma_1$, and 
the fluid is in the resistive regime (section 
\ref{subsec:governing}). This is consistent with 
findings by Wardle (in preparation), that Ohmic conductivity is important at 
this radius for relatively weak fields ($B \lesssim 100$ mG) when the density 
is sufficiently high (e.g. $n_H \gtrsim 10^{12} \percc$ for $B = 1$ mG). 
Both components of the 
conductivity tensor perpendicular to the field increase with 
height initially, reflecting the enhanced ionisation fraction at higher 
$z$.  There is then a 
central section of the disc where $\sigma_1 \gg \sigma_2$ (the fluid is in 
the Hall regime) and finally, both terms drop 
abruptly closer to the surface as a consequence of the fall in 
fluid density. Note that the Hall conductivity term 
decreases much more sharply than the ambipolar diffusion component. As a 
result, the ambipolar diffusion term is typically several orders of 
magnitude greater than the Hall term in 
the surface regions of the disc. To reiterate this concept, we repeat here 
that even in regions where $|\sigma_1| \ll \sigma_2$, the growth and 
structure of 
the MRI in the linear regime are still dominated by Hall diffusion if 
$\chi \lesssim |\sigma_1|/\sigma_{\perp}$ (SW03). 

The previous results are in agreement with the notion that 
$|\sigma_1| > \sigma_2$ at intermediate heights, where $|\beta_e| \gg 1$ 
while $\beta_i \ll 1$. At higher 
$z$, typically $|\beta_e| \gg \beta_i \gg 1$ and the fluid is in the 
ambipolar diffusion regime. In these conditions, 
equations (\ref{eq:sig1}) and (\ref{eq:sig2}) show that  
$\sigma_1 \ll \sigma_2$. On the contrary, near the midplane either 
component may dominate, 
depending on the fluid density and the strength of the magnetic field. 
$\sigma_1$ will be larger if the fluid density is relatively small (larger 
radius) and/or the field is stronger. Conversely, if the field is weak and 
the density is sufficiently high, 
$\sigma_{\parallel} \sim \sigma_2 \gg \sigma_1$ and the conductivity will be 
resistive.
This behaviour of the conductivity 
components as a function of height explains, in turn, the dependency of 
$\chi_{crit}$ with $z$: For weak fields it is $\ll 1$ near the midplane 
($\sigma_2 \gg \sigma_1$), but if 
$B$ is sufficiently strong, $\chi_{crit}$ close to the midplane $\sim 1$. It 
then increases with $z$ and levels in the 
intermediate sections of the disc at $\sim 1$ ($\sigma_1 \gg \sigma_2$). 
Finally, $\chi_{crit}$ drops sharply near the surface 
($\sigma_2 \gg \sigma_1$ again). 

We now briefly turn our attention to the dependency of the magnetic
coupling with height.  Note that $\chi$ first increases with $z$, as
expected, in response to the enhanced ionisation fraction, and lower
fluid density, away from the midplane.  Closer to the surface, it
decreases again, as a result of the abrupt drop in the magnitudes of
the conductivity components.  In the following subsections we explore
more fully the dependency of $\chi$ versus $\chi_{crit}$ with height,
which indicates which non-ideal MHD regime is locally dominant, for
the disc models of interest.

\subsubsection{Minimum-mass solar nebula disc}
    \label{subsubsec:solar_nebula}
 
Figs.  \ref{fig:1AUion} to \ref{fig:10AUion} present curves of $\chi$
(solid lines) and $\chi_{crit} \equiv |\sigma_1|/\sigma_{\perp}$
(dashed lines) as a function of height for the fiducial model.
Results are shown for $R = 1$, $5$ and $10$ AU and different choices
of the magnetic field strength.  Cosmic rays are either included (top
panel of each figure) or excluded from the disc (bottom panels).
Bottom (solid) and leftmost (dashed) curves correspond to $B = 1 mG$
in all cases.  $B$ changes by a factor of $10$ between curves, except
that for $5$ and $10$ AU the top (and rightmost) curves correspond to
the maximum field strength for which perturbations grow.

We note that in all plots, the magnetic coupling near the midplane
increases with $B$.  In these regions of the disc, both $\beta_i$ and
$\beta_e$ are typically $\ll 1$ (except possibly when $B$ is very
strong) and from equation (\ref{eq:sigperp}),

\begin{equation}
	\label{eq:sigperp_1}
\sigma_{\perp} \approx \frac{c e n_e}{B}(\beta_i - \beta_e) \,,
\end{equation}

\noindent
which is independent of $B$. As a result, $\chi \propto B^2$. On the 
contrary, near the surface the $\beta_j$ can be very large and

\begin{equation}
	\label{eq:sigperp_2}
\sigma_{\perp} \approx \frac{c e n_e}{B} \frac{(\beta_i - \beta_e)}
	{\beta_e \beta_i} \propto \frac{1}{B^2} \,.
\end{equation} 

\noindent
Because of this, near the surface $\chi$ is quite insensitive to changes in 
$B$, as evidenced in Figs. \ref{fig:1AUion} to \ref{fig:10AUion}. 

Note that when cosmic rays are assumed to be excluded from the disc,
there is a discontinuity in the curve of $\chi$ vs.  $z/H$ at $R = 1$
and $5$ AU. This discontinuity is caused by the drop in the ionisation
fraction at the height where x-rays are not able to penetrate any
further within the disc (see bottom panels of Figs.  \ref{fig:1AUion}
and \ref{fig:5AUion}).  It is, therefore, not present at $10$ AU,
because x-rays reach the midplane at this radius (see bottom panel of
Fig.  \ref{fig:10AUion}).

In general, when $B$ increases, the region around the midplane where
Hall diffusion is locally dominant ($\chi <
|\sigma_1|/\sigma_{\perp}$, SW03) is reduced, as a result of the
stronger magnetic coupling.  At $1$ AU, when cosmic ray ionisation is
included, this criterion is satisfied in the inner sections of the
disc for all magnetic field strengths of interest (Fig.
\ref{fig:1AUion}, top panel).  Above this region, there is a
relatively small section where ambipolar diffusion is locally dominant
while near the surface, ideal-MHD holds for all magnetic field
strengths.
 
For $R = 5$ AU (Fig.  \ref{fig:5AUion}, top panel), Hall diffusion
dominates for $z/H \lesssim 2.2$ when $B \lesssim 10$ mG, but for $10$
mG $\lesssim B \lesssim 100$ mG, ambipolar diffusion is locally
dominant in the inner sections of the disc ($|\sigma_1|/\sigma_{\perp}
< \chi < 10$).  For stronger fields, the magnetic coupling is such
that $\chi > 10$ even at the midplane and the fluid is in ideal-MHD
conditions over the entire cross-section of the disc.  At this radius,
the fluid is in the resistive regime only for very weak fields ($B
\lesssim 1$ mG) and very high densities ($n_H > 10^{12} \percc$;
Wardle, in prep.).  Similarly, results at $R = 10$ AU (Fig.
\ref{fig:10AUion}, top panel), show that ambipolar diffusion dominates
in the inner sections of the disc for $B < 10$ mG. For stronger
fields, ideal-MHD is a good aproximation at all $z$.

When cosmic rays are assumed to be excluded from the disc, the
previous results at $1$ AU (bottom panel of Fig.  \ref{fig:1AUion})
are largely unchanged, except for those obtained with $B = 1$ G, where
Hall diffusion dominates now for $z/H \sim 1.8$, up from $\sim 1$ in
the previous case.  This is due to the sharp fall in the magnetic
coupling in the region which x-rays are unable to reach.  At $5$ AU,
the Hall regime is now dominant near the midplane for all studied
magnetic field strengths, while for $R = 10$ AU there is now a Hall
dominated central region for $B \lesssim 10$ mG. It extends to $z/H
\sim 1.3$.  When $B \lesssim 100$ mG, ambipolar diffusion dominates
near the midplane, but for stronger fields ideal MHD holds for all
$z$.

Finally, we observe that (as expected), the magnetic coupling at the
midplane increases with radius in response to the higher ionisation
fraction in the central sections of the disc at larger radii.  In
fact, $\chi_o$ increases by $3$ - $4$ orders of magnitude between $R =
1$ and $10$ AU. On the contrary, the coupling at the surface does not
change as much with radius, decreasing only from $\sim 22$ to $\sim
2.5$ between the same radii.

\subsubsection{A more massive disc}
     \label{subsubsec:thicker}

Fig. \ref{fig:ionthick} displays curves of $\chi$ and $\chi_{crit}$ as a 
function of height for a more massive disc, as detailed in section 
\ref{subsec:model}. The ionisation balance is calculated assuming that 
cosmic rays penetrate the disc and results are shown for $R = 1$, $5$ and 
$10$ AU for the same range of magnetic field strengths explored in the 
minimum-mass solar nebula model. 

Increasing the surface density causes the ionisation fraction near the
midplane to drop sharply.  As a result, the magnetic coupling is
drastically reduced in these regions at all radii.  This is especially
noticeable at $1$ AU, where in this case x-rays are completely
attenuated for $z/H < 2.6$.  This, together with a very weak
ionisation rate ($\zeta_{CR} \sim 10^{-18}$ for $Z/H \approx 2.6$ and
it is negligible at $z = 0$), causes $\chi$ to be very low --and
fairly independent of $z$-- in this section of the disc.

The weaker coupling at low $z$ results in Hall diffusion being locally
dominant over a larger cross-section of the disc in this model.  We
find that for $R = 1$ AU, there is again a Hall dominated region about
the midplane for all studied $B$.  This section now extends to $z/H
\sim 2.3$ for $B = 1$ G, up from $z/H \sim 1$ in the minimum-mass
solar nebula disc.  For weaker fields, these regions are also larger.

For $R = 5$ AU, Hall diffusion is now dominant near the midplane for
$B < 1$ G. For comparison, in the previous case Hall conductivity was
locally important for $B < 10$ mG. Furthermore, ambipolar diffusion
dominates over a small cross-section of the disc for all but the
strongest ($B \gtrsim 1$ G) magnetic field strengths at this radius.
This contrasts with the previous case, where ideal MHD held at all $z$
when $B \gtrsim 100$ mG. Even at $10$ AU, Hall diffusion now dominates
near the midplane when $B \lesssim 10$ mG. When $B$ is stronger than
this, but still $\lesssim 100$ mG, ambipolar diffusion predominates
for $z/H \lesssim 2$, while in the minimum-mass solar nebula,
ideal-MHD held for $B \gtrsim 10$ mG.
 
In the next section we analyse the effect of different conductivity
regimes being dominant at different heights, in the structure and
growth rate of MRI perturbations.  Different disc models, radial
locations and magnetic field strengths are discussed.

\begin{figure} 
\centerline{\epsfxsize=8cm \epsfbox{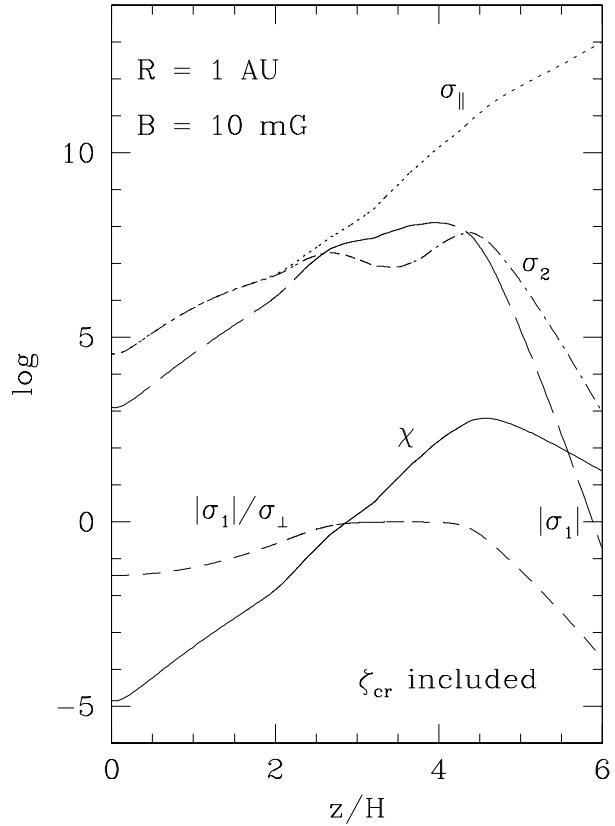}}\vskip 0cm
\caption{Example of the dependency of the conductivity components
parallel ($\sigma_{\parallel}$) and perpendicular $\sigma_1$ and
$\sigma_2$) to the field, magnetic coupling $\chi$ and $\chi_{crit}
\equiv |\sigma_1|/\sigma_{\perp}$ as a function of height.  Results
shown correspond to the fiducial model for $R = 1$ AU and $B = 10$ mG.
Note that for $z/H \lesssim 2$, $\sigma_{\parallel} \sim \sigma_2 \gg
\sigma_1$ and the conductivity is resistive.  There is then a central
section for which $|\sigma_1| \gg \sigma_2$ (the fluid is in the Hall
conductivity regime), while for higher vertical locations
$\sigma_{\parallel} \gg \sigma_2 \gg \sigma_1$ and ambipolar diffusion
dominates.  In the region where $\chi < \chi_{crit}$, Hall diffusion
still dominates the structure and growth rate of the MRI (SW03).}
\label{fig:ionexample}
\end{figure}

\begin{figure} 
\centerline{\epsfxsize=7.5cm \epsfbox{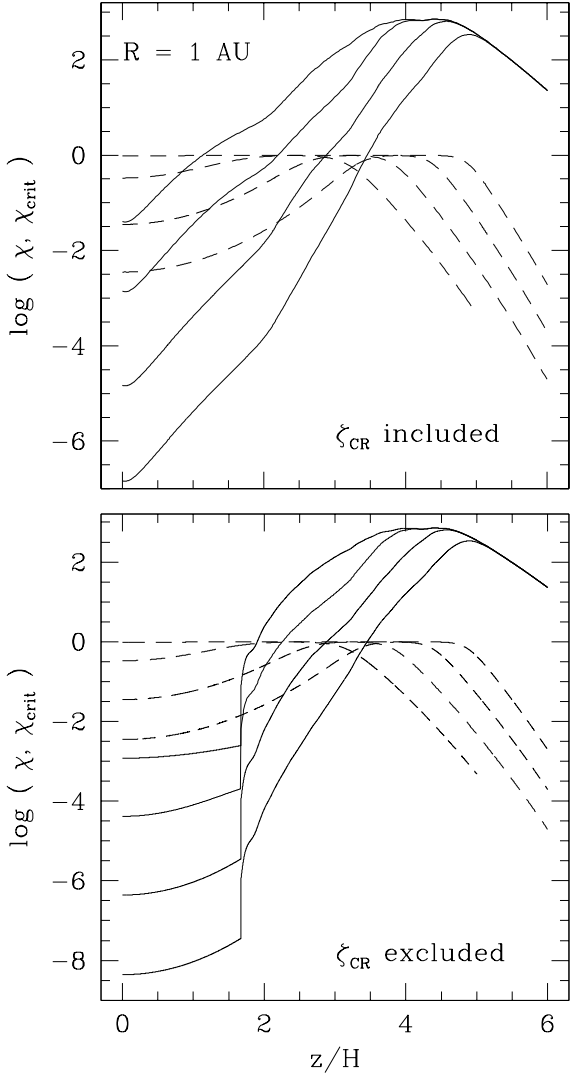}}\vskip 0cm
\caption{Comparison of the local magnetic coupling $\chi$ (solid
lines) and $\chi_{crit} \equiv |\sigma_1|/\sigma_{\perp}$ (dashed
lines) for the fiducial model at $R = 1$ AU and different choices of
the magnetic field strength.  In each case, Hall diffusion is dominant
in the regions where $\chi < \chi_{crit}$ (SW03).  Ambipolar diffusion
dominates where $\chi_{crit} < \chi \lesssim 10$.  When $\chi$ is
stronger than this, the fluid is in nearly ideal-MHD conditions (W99).
From top to bottom (solid lines), and right to left (dashed lines),
the magnetic field drops from $1$ G to $1$ mG. $B$ changes by a factor
of $10$ between curves in all cases.  Top panel: Cosmic rays are
present.  Hall diffusion dominates for $z/H \lesssim 3.5$ to $z/H
\lesssim 1$ for $B$ increasing from $1$ mG to $1$ G, respectively.
Bottom panel: Cosmic rays are excluded from the disc by protostellar
winds.  Note the discontinuity in $\chi$ at the height below which
x-rays are completely attenuated.}
\label{fig:1AUion}
\end{figure}

\begin{figure} 
\centerline{\epsfxsize=7.5cm \epsfbox{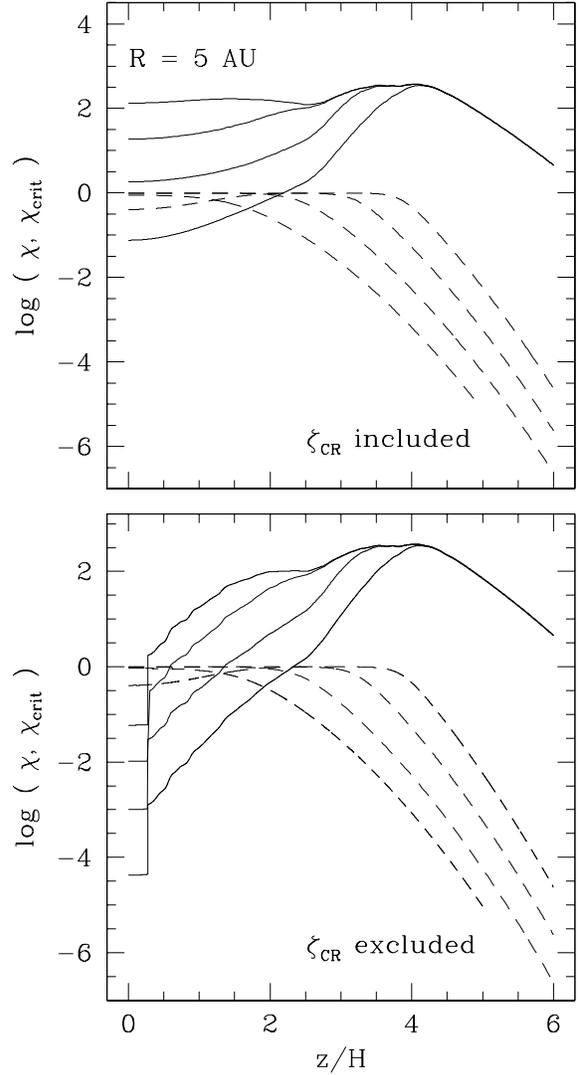}}\vskip 0cm
\caption{As per Fig. \ref{fig:1AUion} for $R = 5$ AU. Top (and rightmost) 
lines correspond here to $B = 795$ mG (top panel) and $B = 615$ mG (bottom 
panel), which are the maximum field strengths for which perturbations grow in 
each scenario. When cosmic 
rays are included, the Hall regime dominates near the midplane for 
$B \lesssim 10$
mG. Similarly, for $B < 100$ mG, there is an intermediate region where 
$\sigma_1/\sigma_{\perp} < \chi < 10$, and ambipolar diffusion is
dominant. For stronger fields, ideal-MHD holds for all $z$. When cosmic 
rays are excluded, Hall diffusion dominates near the midplane, and ambipolar 
diffusion is dominant at intermediate heights, for all $B$ of interest here.}
\label{fig:5AUion}
\end{figure}

\begin{figure} 
\centerline{\epsfxsize=7.5cm \epsfbox{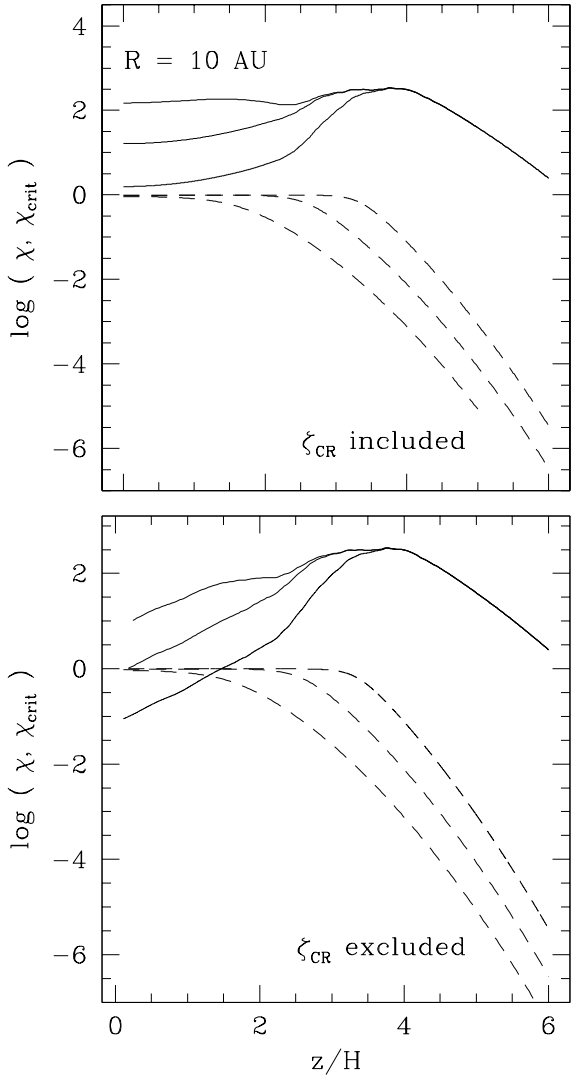}}\vskip 0cm
\caption{As per Figs. \ref{fig:1AUion} and \ref{fig:5AUion} for $R = 10$ AU. 
From top to bottom (and right to left), lines 
correspond here to $B = 100$, $10$ and $1$ mG. 
In this case, if cosmic 
rays are present, Hall diffusion is not dominant locally for any field 
strength, given the strong 
magnetic coupling near the midplane. Ambipolar diffusion 
dominates when $B \lesssim 10$ mG, but for stronger fields 
ideal-MHD holds over the entire cross-section of the disc. If cosmic rays are 
removed, the Hall regime is relevant near the midplane only for 
$B \lesssim 10$ mG. When $B \gtrsim 100$ mG, ideal MHD holds for all $z$.}
\label{fig:10AUion}
\end{figure}

\begin{figure} 
\centerline{\epsfxsize=7.5cm \epsfbox{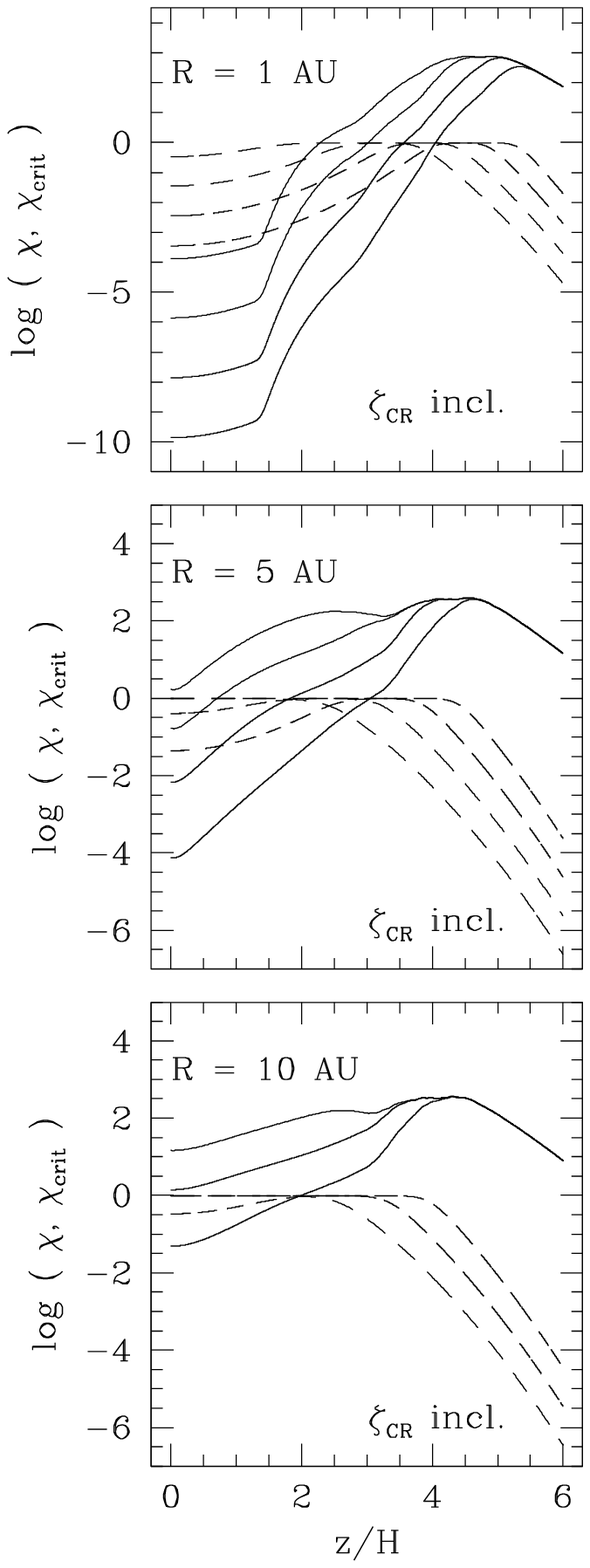}}\vskip 0cm
\caption{As per Figs. \ref{fig:1AUion} to \ref{fig:10AUion} for $R = 1$, $5$ 
and $10$ AU for a more massive disc, such that $\Sigma_o^{'} = 10\Sigma_o$ and 
$\rho_o^{'} = 10\rho_o$. 
Cosmic rays are present. From top to bottom (and right to left), lines 
correspond to $B = 1$, 
$0.1$, $10^{-2}$ and $10^{-3}$ G ($R = 1$ and $5$ AU), and $B = 0.1$, 
$10^{-2}$ and $10^{-3}$ G ($R = 10$ AU). The increased surface density causes 
the magnetic coupling to drop near the midplane. As a result, Hall 
conductivity dominates over a larger section of the disc around the midplane, 
while ambipolar diffusion is important (at higher $z$) for stronger fields 
than it was the case in the minimum-mass solar nebula model.}
\label{fig:ionthick}
\end{figure}

\section{MAGNETOROTATIONAL INSTABILITY}
\label{sec:results2}

\subsection{Structure of the perturbations}
\label{subsec:structure}

\subsubsection{All unstable modes at 1 AU}
\label{subsubsec:allmodes}

\begin{figure*} 
\centerline{\epsfxsize=17cm \epsfbox{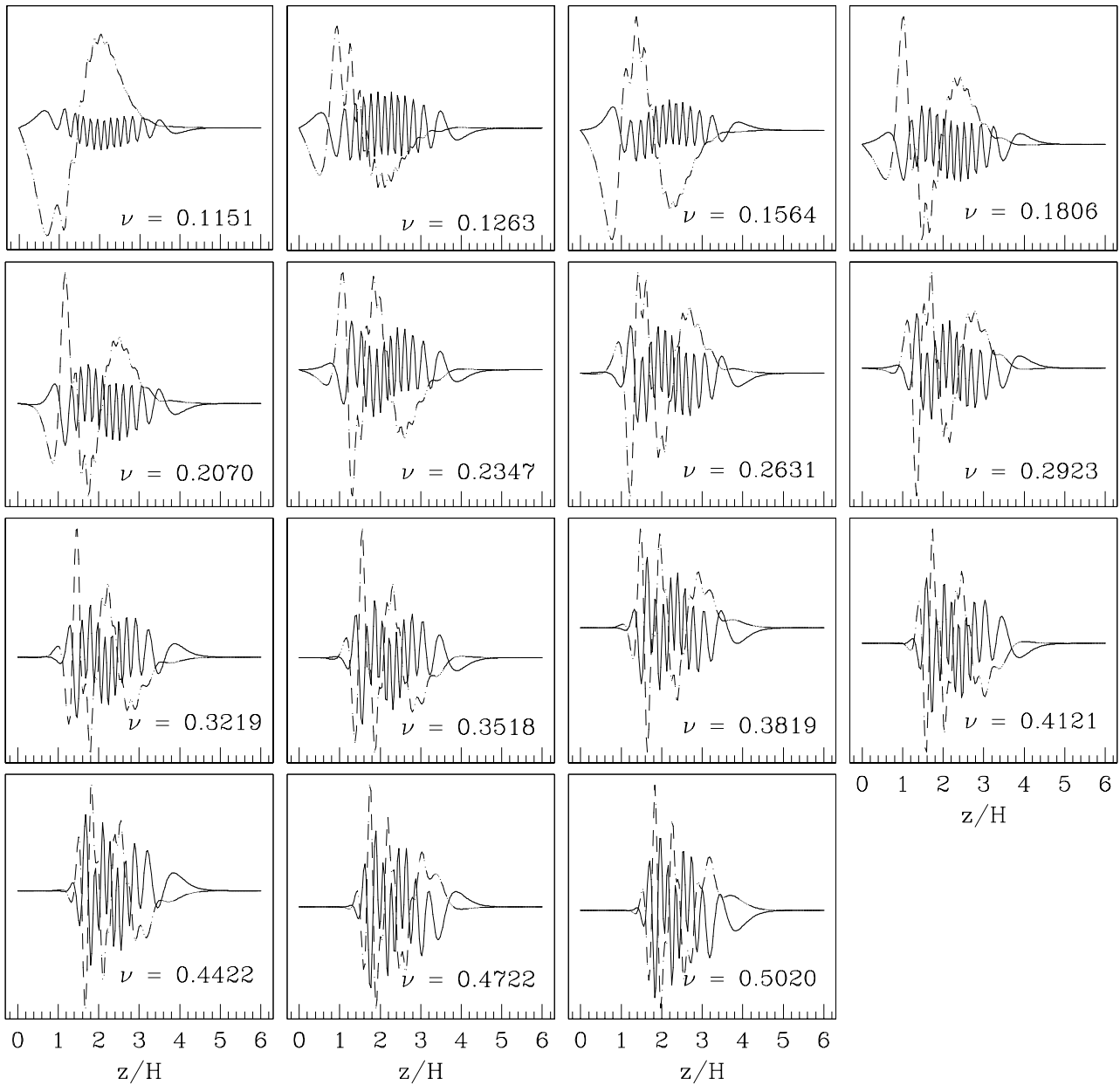}}\vskip 0cm
\caption{Structure of all unstable modes for the fiducial model at $R = 1$ 
AU and $B = 100$ mG. In each case solid lines show $\delta B_r$ while dashed 
ones correspond to $\delta B_{\phi}$. The 
growth rate is indicated in the lower right corner of each panel. There 
are $15$ unstable
modes, with $0.1151 \leq \nu \leq 0.5020$. Some perturbations grow at the 
midplane, particularly when $\nu < 0.1806$. Modes growing faster than this 
show a dead region around the midplane, whose vertical extent increases with 
the growth rate 
of the perturbations. Note the asymmetry of $\delta B_r$ and 
$\delta B_{\phi}$ about zero, especially in slowly growing modes.}
\label{fig:1AUmodes}
\end{figure*}

Fig. \ref{fig:1AUmodes} shows the structure and growth rates of all unstable 
MRI perturbations 
for the fiducial model at $1$ AU and $B = 100$ mG. Solid lines denote 
$\delta B_r$ while dashed 
lines correspond to $\delta B_{\phi}$. This notation is observed in this 
paper in all plots that display the structure of the instability.

The fastest growing perturbation in this case grows at $\nu = 0.5020$ and 
there are $15$ unstable modes with 
$0.1151 \leq \nu \leq 0.5020$. Slow growing perturbations, with 
$\nu < 0.1806$, are 
active even at the midplane, while faster modes exhibit a central magnetically 
inactive (dead) zone (Gammie 1996, Wardle 1997). 
Moreover, the extent of the dead zone increases with the growth rate and for 
the fastest growing mode it extends to $z/H \sim 1.6$. 
We observe that $\delta B_r$ and $\delta B_{\phi}$, specially in the slow 
growing 
modes, are asymmetrical about zero and their averages over vertical 
sections of the disc have opposite signs. 
This appears to be related to the dependency of these two fluid variables 
in equation (\ref{eq:induc_phi_final}). 

\subsubsection{Most unstable modes at different radii, including cosmic ray 
ionisation}
\label{subsubsec:numax}

\begin{figure*} 
\centerline{\epsfxsize=12cm \epsfbox{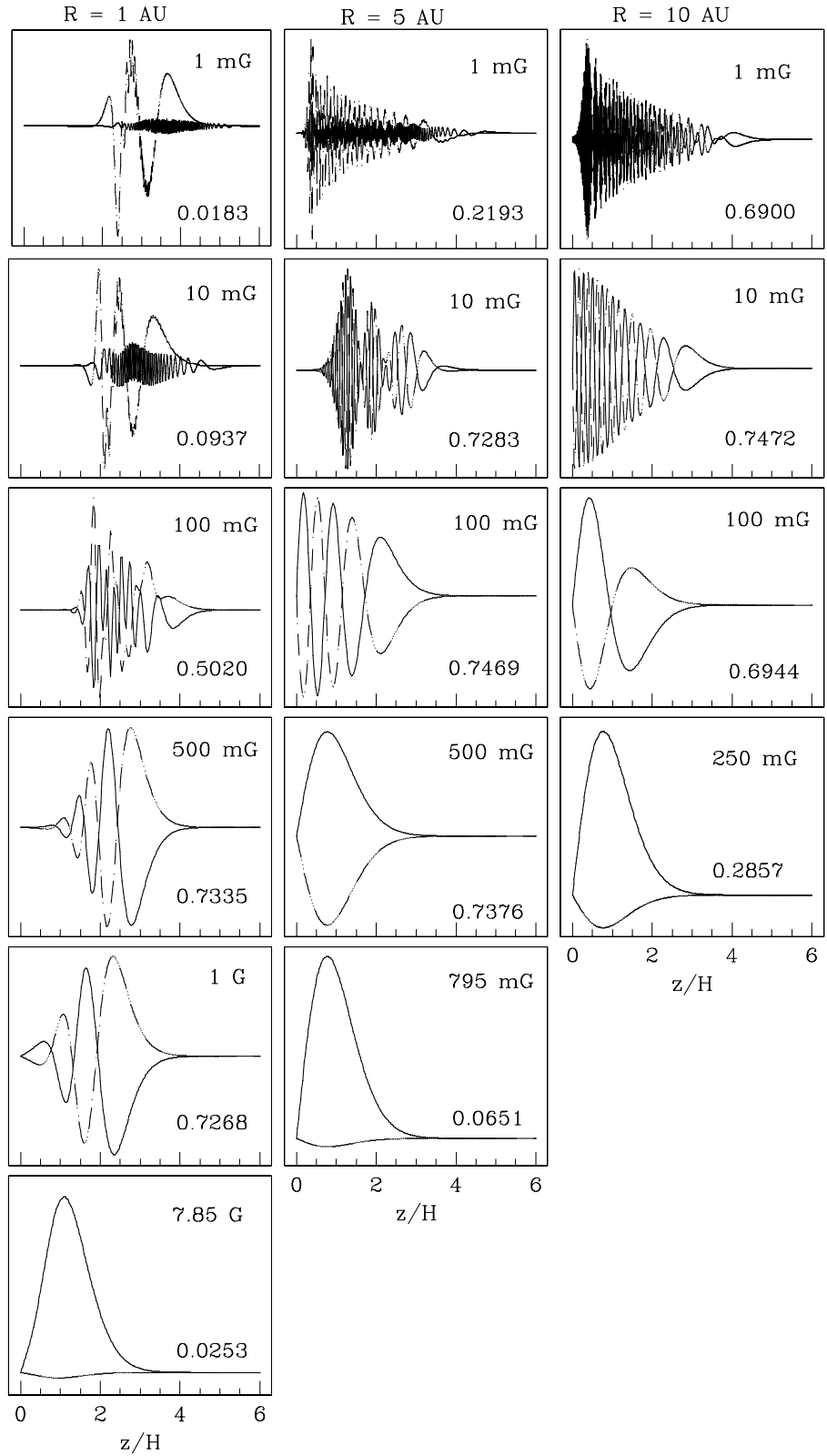}}\vskip 0cm
\caption{Comparison of the structure of the most unstable modes for the 
fiducial model at $R = 1$, $5$ and $10$ AU for different choices of the 
magnetic field strength. $B$ 
is indicated in the upper right hand of each panel while the 
growth rate is shown in the bottom right hand side. Results are shown for 
$B$ spanning from $1$ mG, the weakest magnetic field strength for which 
unstable modes 
could be computed, to the maximum strength for which unstable modes grow, 
which changes with the radius. Note that for all three radii, the wavelength 
of the perturbations 
increases with the strength of the magnetic field, as expected. At $1$ AU, for 
$B < 500$ mG, 
there is a central dead zone which extends to $z/H \lesssim 2$. At $5$ AU, 
only 
modes for $B < 100$ mG exhibit a dead region. At this radius the magnetic 
coupling, even at the midplane, is sufficiently large for ideal MHD to be a 
good approximation for stronger fields. Finally, for $R = 10$ AU, ideal 
MHD 
holds throughout the cross-section of the disc for $B \gtrsim 10$ mG and 
perturbations in this region of parameter space grow even at the midplane.}
\label{fig:allmodes1}
\end{figure*}

The weakest strength for which unstable modes could be computed was $1$ mG. 
This is a computational, rather than a physical limit: the dependency of 
$\nu_{max}$ with $B$ in the weak-field limit, shows no evidence for a minimum 
required field strength for the instability to grow (see section 
\ref{subsubsec:grow_cond}).
A comparison of the structure of the most unstable modes for the fiducial 
model at $R = 1$, $5$ and $10$ AU is shown in Fig. \ref{fig:allmodes1}. We 
display solutions for $B$ from $1$ mG, up to the maximum field strength for 
which unstable modes grow for each radius. Unless stated otherwise, this 
criterion is followed in all 
plots that show the structure of the fastest growing modes as a function of 
$B$ in this study. Note that the maximum value of $B$ is dependent on the 
disc model, the ionising agents incorporated and radius of interest. This 
does not imply that $1$ mG is necessarily the weakest field strength for which 
unstable modes exist. In fact, from 

Both the structure and growth rate of these perturbations are shaped 
by the competing action of different conductivity components, whose 
relative importance change with height. We defer the analysis of their growth 
rate to section \ref{subsec:growth}. Here, we discuss the structure as 
a function of $z$. Note first of all that at each radius, the wavelength of 
the perturbations increase with the
magnetic field strength, as expected by both ideal-MHD and non-ideal MHD 
local analyses (Balbus \& Hawley 1991, W99).  

At $1$ AU (leftmost 
column of Fig. \ref{fig:allmodes1}), for $B \lesssim 500$ mG, the region 
next to the midplane is a 
magnetically inactive zone (Gammie 1996, Wardle 1997). The extent of 
this region decreases as the field gets stronger. For example, for 
$B = 1$ mG it extends from the 
midplane to $z/H \sim 1.8$, but
when $B > 500$ mG, there is no appreciable dead region, a result of the 
stronger magnetic coupling close to the midplane and the relatively small
wavenumber of the perturbations.
At this radius, when $B$ is relatively strong ($B > 500$ mG), the 
amplitude of these modes increase with $z$, a property that is 
typical of MRI perturbations driven 
by ambipolar diffusion. Ambipolar diffusion modes have this property because, 
as the local analysis (W99) 
indicates, in this limit
the local growth of unstable modes increases with the magnetic coupling, 
which (except in the surface regions) increases with $z$ (see Fig. 
\ref{fig:ionexample}). As a result, the 
local growth rate of the MRI also increases with height in this regime and is 
able to drive the amplitude of global unstable modes to increase.  
This explains the shape of the envelope of these modes. Finally, note that 
when the magnetic field
is weak, the perturbations' wavenumber is very high, and $\delta B_r$ and 
$\delta B_{\phi}$ are not symmetrical about $z = 0$. 

At $5$ AU, when $B \gtrsim 100$ mG, perturbations grow even at the midplane. 
For weaker fields, they exhibit only a very small 
dead zone which extends to $z/H \lesssim 0.5$. Note also that the envelope of 
these modes, particularly for 
$B \gtrsim 100$ mG, is fairly flat. This is explained by recalling that the 
magnetic coupling at the midplane in this region is very 
high ($\chi \sim 90$ for $B = 500$ mG, Fig. \ref{fig:5AUion}, top panel), so 
the ideal-MHD approximation holds. Under
these conditions, unstable modes peak at the node closest to the midplane, 
given that the local growth rate is not a 
strong function of $\chi$ and does not vary by much with height (see 
also top row of Fig. 6 of SW03, which shows similar envelopes for 
perturbations obtained with $\chi_o = 100$ and 
different configurations of the conductivity tensor). On the other hand, for 
$B \lesssim 100$ mG, non-ideal MHD effects are important. When $B = 10$ mG, 
ambipolar diffusion is dominant for $z/H \lesssim 2.3$. This conductivity term
is likely to
drive this perturbation's structure, as evidenced by the central dead zone 
and the envelope peaking at an intermediate height. For 
$B = 1$ mG, Hall conductivity is dominant close to the midplane 
($z/H \lesssim 2.2$), the region where the envelope of this perturbation 
peaks. Furthermore, the high wavenumber suggests that its structure is 
determined by local effects. As a result, Hall conductivity is likely to 
drive the structure of this mode.

Finally, for $R = 10$ AU there is no appreciable dead zone 
for $B \gtrsim 10$ mG, given the strong magnetic coupling at this radius. Even 
for $B = 1$ mG, $\chi_o \sim 3$, a
figure that increases to $\sim 100$ for $B = 100$ mG. 
As a result, the fluid is in nearly ideal-MHD conditions (see section 
\ref{subsubsec:solar_nebula}), 
which explains the flat envelope of these modes. 

\subsubsection{Most unstable modes at different radii, excluding cosmic ray 
ionisation}
\label{subsubsec:numax2}

\begin{figure*} 
\centerline{\epsfxsize=12cm \epsfbox{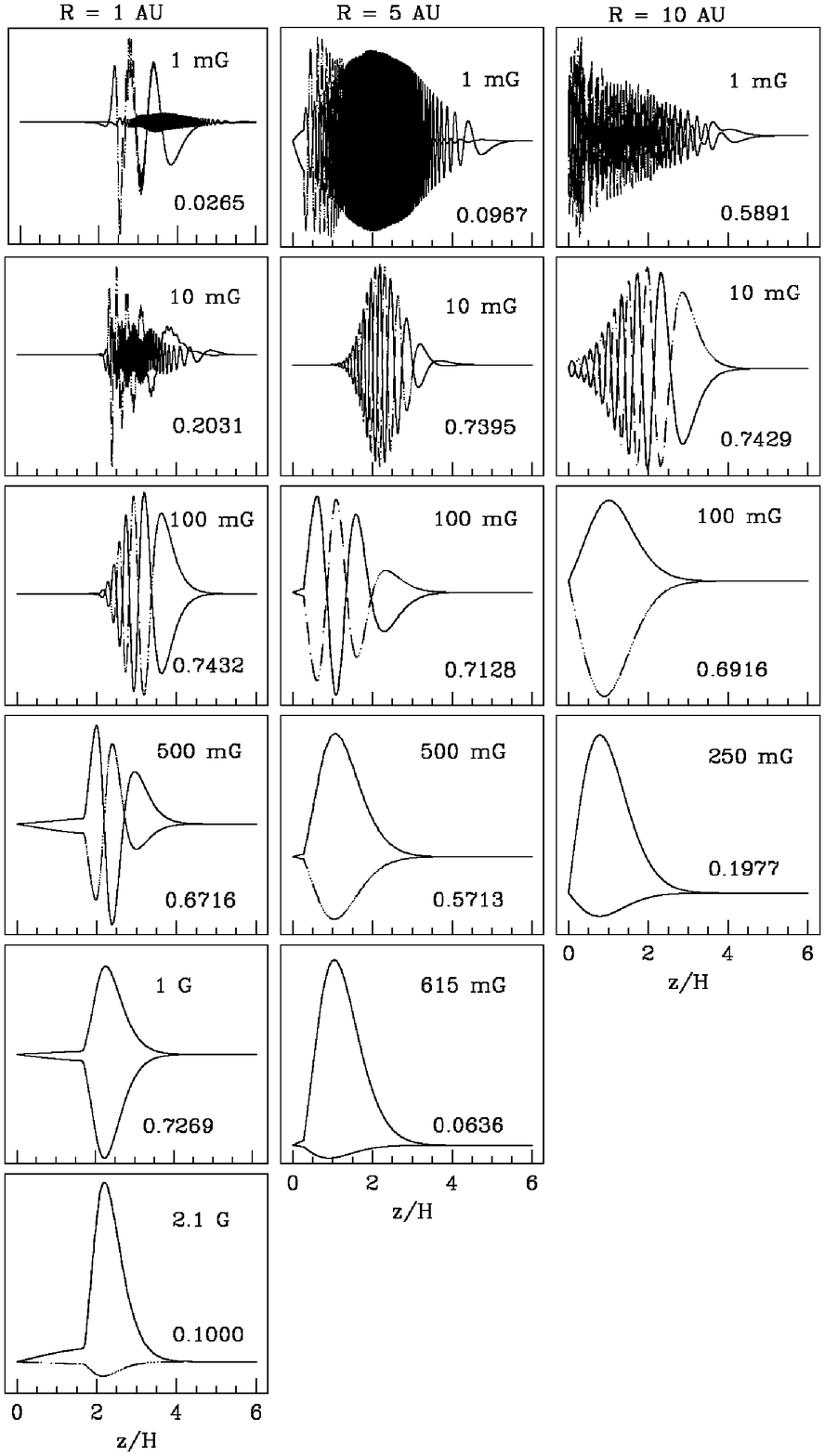}}\vskip 0cm
\caption{As per Fig. \ref{fig:allmodes1} assuming cosmic rays are excluded 
from the disc. Note the increased extent of the central dead zone at $1$ AU 
($B \lesssim 100$ mG) and the kink in the amplitude of the perturbations for 
stronger fields. This feature is caused by the sharp change in the ionisation
balance of the disc at the height where x-ray ionisation becomes active. As 
expected, this effect is less pronounced at $5$ AU and unnoticed at $10$ AU, 
where x-rays reach the midplane.}
\label{fig:allmodes2}
\end{figure*}

We also calculated the structure of the most unstable modes at the same three 
radial positions under the assumption that cosmic rays are excluded from the 
disc by protostellar winds (Fig. \ref{fig:allmodes2}). In this case the 
central dead zones at $1$ AU ($B \lesssim 100$ mG) extend over a larger 
cross-section of the disc, 
given that cosmic rays (when present) are the main source of 
ionisation near the 
midplane at this radius (see section \ref{subsec:ionrate}). Without them, the
electron fraction below $z/H \sim 1.7$ (where x-rays are 
completely attenuated) plummets, causing the amplitude of MRI
perturbations in this section of the disc to be severely reduced. As a 
result, when $B \lesssim 100$ mG, perturbations are damped for 
$z/H \lesssim 2$. Another difference with the previous case is 
noticeable for stronger fields ($B \geq 500$ mG), where before the MRI was 
active even at the midplane. In this case, the abrupt change in the 
ionisation balance at the height where x-ray ionisation becomes active, 
(Fig. \ref{fig:ionbalance}, top panel), causes the current to be 
effectively discontinuous there and produces the observed kink in the 
amplitude of these modes. 

In 
the zone where x-rays are excluded ($z/H \lesssim 1.7$), the main ionising 
agent is the decay of radioactive elements within the disc. However, they 
can only produce very weak magnetic coupling. For example, $\chi_o 
\sim 10^{-8}$ for $B = 1$ mG, and it 
only increases to $\sim 10^{-3}$ when $B = 1$ G. As a result, the amplitude 
of the perturbations near the midplane, even for strong fields 
($B \gtrsim 500$ mG), is very small at this 
radius. If the abundance of metals in the gas phase is reduced 
from the fiducial value adopted here ($\delta_2 \approx 0.02$) to (say)  
$ \sim 2 \ee -3 $, the zone ionised 
only by radioactivity becomes magnetically dead (as expected). The 
perturbations' growth is only marginally affected.

For $R = 5$ AU, we also observe a small
kink in the perturbations' amplitude for $B \gtrsim 100$ mG. In this case
it occurs much closer to the midplane, because at this radius 
x-rays penetrate to 
$z/H \sim 0.3$. 
Finally, for $R = 10$ AU, x-rays penetrate the entire cross section of the
disc, so there is no kink in the 
amplitude of these modes. However, the midplane
cosmic ray ionisation rate at this radius is about two orders of magnitude 
larger than that 
of x-rays (in fact they dominate over x-rays for $z/H \lesssim 2.5$, see Fig. 
\ref{fig:ionbalance}, bottom panel),
so excluding them does reduce significantly the magnetic coupling in 
this region. As an illustration of this, note that for $B = 10$ mG, $\chi_o$ 
decreases 
from $\sim 16$ in the previous case to only $\sim 1$ here (see Fig. 
\ref{fig:10AUion}). As a result, for $z/H \lesssim 0.3$ the fluid is 
in the Hall regime while ambipolar 
diffusion dominates at higher $z$. Being a high wavenumber perturbation, the
structure of this mode reflects mainly local fluid conditions. This explains 
the shape of its envelope (see rightmost column of Fig. \ref{fig:allmodes2}, 
second 
panel from the top): flat envelope close to the midplane where the Hall 
effect is dominant and amplitude 
increasing with $z$ at higher vertical locations, driven by ambipolar 
diffusion. On the other hand, for $B = 1$ mG, 
$\chi < |\sigma_1|/\sigma_{\perp}$ for $z/H < 1.3$. This, together with the 
high wavenumber of the perturbations, causes Hall diffusion to shape the 
envelope of this mode.

\subsubsection{Conductivity regime comparison ($\sigma_1B_z > 0$)}
\label{subsubsec:comparison}

\begin{figure*} 
\centerline{\epsfxsize=12cm \epsfbox{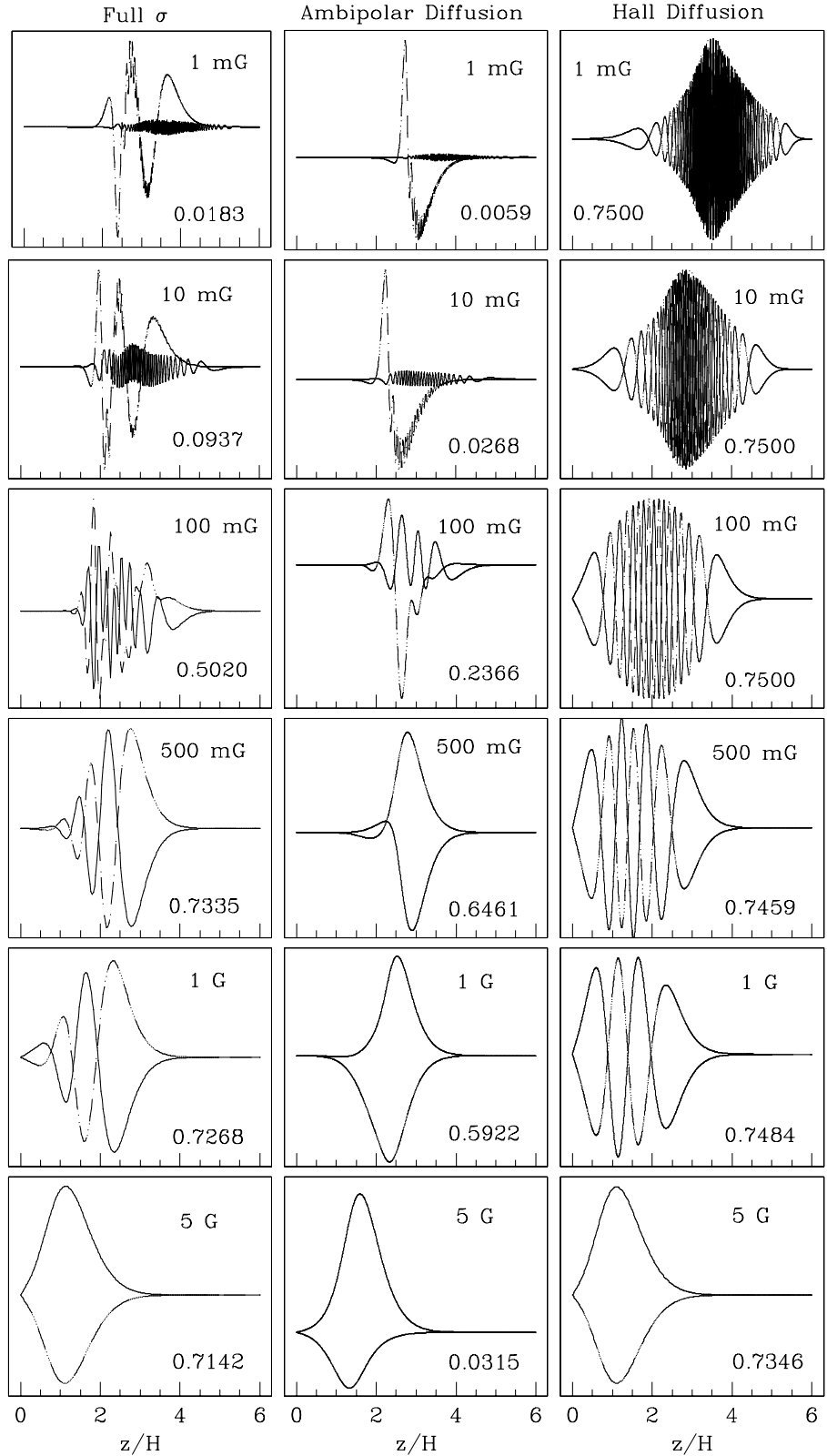}}\vskip 0cm
\caption{Comparison of the structure of the most unstable modes of the MRI 
for the fiducial model at $R = 1$ AU as a function of the magnetic field 
strength for different 
configurations of the conductivity tensor. The field strength spans from $1$ 
mG up to $5$ G, which is the maximum $B$ for which perturbations grow in the 
ambipolar diffusion regime. The left column shows solutions 
obtained with a full conductivity tensor. The middle and right columns 
corresponds  to the 
ambipolar ($\sigma_1 = 0$) and Hall ($\sigma_2 = 0$, 
$\sigma_1B_z > 0$) 
approximations, respectively. Note that when the magnetic field is weak 
($B \lesssim 100$ mG), full conductivity perturbations have a higher 
wavenumber, and 
grow closer to the midplane, than modes in the ambipolar diffusion limit. This 
reflects the contribution of the Hall effect. For stronger fields, the shape 
of the envelope appears to be driven by ambipolar diffusion (amplitude 
increases with height), with the Hall term increasing the perturbations' 
wavenumber.}
\label{fig:regimecomp}
\end{figure*}

It is interesting to compare the structure of the most unstable perturbations 
obtained with different configurations of the conductivity tensor (assuming 
that different conductivity regimes dominate over the entire 
cross section of the disc) against the full conductivity results discussed in
section \ref{subsubsec:numax}. This way we can explore more fully the 
effects of different 
conductivity components in the overall properties of the instability. Fig. 
\ref{fig:regimecomp}  presents such solutions for the fiducial model at 
$R = 1$ AU as a function of the strength of the 
magnetic field. The left column shows solutions computed with a full 
conductivity 
tensor, while the middle and right ones display modes obtained using the 
ambipolar diffusion ($\sigma_1 = 0$) and Hall regime ($\sigma_2 = 0$, 
$\sigma_1B_z > 0$) approximations, respectively. 

Note that, for all magnetic field strengths shown here, perturbations computed 
using the ambipolar diffusion approximation have a more extended 
central dead zone than modes incorporating a full conductivity tensor. 
This is in agreement with the local analysis (W99), which showed that 
the growth of MRI perturbations in the ambipolar diffusion limit decreases 
steadily 
with the local magnetic coupling. 
As a result, the local growth of these modes is 
severely restricted near the midplane. This is 
especially effective for weaker fields, when global effects are less 
important, due to the high wavenumber of the perturbations. For example, 
when $B = 1$ mG, the magnetically dead zone in modes found using the 
ambipolar diffusion approximation extends to $z/H \sim 2$. The thickness of 
the dead region 
decreases for stronger $B$, but even with $B = 1$ G, there is a small section 
($z/H \lesssim 0.5$) in which perturbations do not grow. 
 
Turning now our attention to Hall perturbations, we observe that their 
amplitude is fairly stable, especially for strong fields ($B \gtrsim 500$ mG), 
where they show the characteristic flat envelope reported by SW03. When the 
field is weaker than this, perturbations  are
damped near the midplane. This is particularly evident for $B < 100$ mG and is
probably related to the fact that at $1$ AU the magnetic coupling 
at the midplane in this limit is very low. Even for $B = 100$ mG, $\chi_o$ in 
the Hall 
limit $\sim 10^{-4}$ and it decreases to $\sim 10^{-10}$ for $B = 1$ mG (note 
that these values are significantly smaller than the ones shown in Fig. 
\ref{fig:1AUion}, which were obtained with a full conductivity tensor). Local
results highlight that
the maximum local growth rate of MRI perturbations in the Hall limit is 
effectively unchanged from the ideal case when $\chi \rightarrow 0$ (W99). 
However, the global analysis presented here shows that when the coupling is 
very low, the amplitude of global Hall limit perturbations can be
damped close to the midplane (see rightmost column of Fig. 
\ref{fig:regimecomp}, top three panels). On the other hand,
there is only a small dead zone for $B \sim 1$ mG, which extends from the 
midplane to $z/H \sim 0.5$. For stronger fields, there is no appreciable dead 
region.

In general, the structure of full $\bmath{\sigma}$ perturbations reflect the 
contribution 
of Hall as well as ambipolar diffusion conductivity terms. When the magnetic 
field is strong ($B > 100$ mG), ambipolar 
diffusion is locally dominant over a more extended section of the disc, and 
the dead zone of perturbations in this limit is much smaller than they are 
for weaker fields. On 
the other hand, modes in the Hall limit grow now even at the midplane and 
have a significantly
higher wavenumber than ambipolar diffusion perturbations. This is reflected 
in the structure of full $\bmath{\sigma}$ modes in this region of parameter 
space: Their envelope is shaped by 
ambipolar diffusion (the amplitude increases with height), but the wavenumber
is higher and they grow closer to the midplane than pure ambipolar diffusion 
modes do. This 
 reflects the contribution of the Hall effect. These results are also in 
agreement with similar trends found with illustrative calculations in SW03. 

It was discussed in section \ref{subsubsec:regimes} that at this radius ($1$ 
AU), Hall 
conductivity is locally dominant near the midplane for all $B$ of interest. 
The transition to the 
zone where ambipolar diffusion dominates occurs at a lower $z$ for stronger 
fields (Fig. \ref{fig:1AUion}, top panel). This analysis reveals that the Hall 
effect modifies the structure of MRI modes even when Hall diffusion is 
locally dominant only for a small section close to the midplane of the disc.
In section \ref{subsubsec:grow_cond} it will be discussed how the Hall effect 
also alters the growth of all fastest growing modes at this radius. 

At $5$ AU, ambipolar diffusion is important in the inner sections of the disc 
for $B \lesssim 100$ mG (Fig. \ref{fig:5AUion}). It is expected that 
perturbations obtained using this approximation will be different from the 
corresponding full $\bmath{\sigma}$ modes in this region of parameter space. 
Fig. \ref{fig:AUcond3} compares solutions obtained with different 
configurations of the conductivity tensor at this radius. Ambipolar diffusion 
modes (left column) are indeed different from full $\bmath{\sigma}$ 
($\sigma_1B_z > 0$) ones (middle column) when $B \lesssim 100$ mG. For 
stronger fields, ideal-MHD holds, so the structure of MRI modes computed using 
different configurations of the conductivity tensor are alike, as expected.

\subsubsection{Conductivity regime comparison ($\sigma_1B_z < 0$)}
\label{subsubsec:sigma_1}

So far we have discussed solutions obtained with a $\sigma_1B_z >0$ Hall 
conductivity term, which corresponds to the case where the magnetic field and 
angular velocity vectors of the disc are parallel 
($\bmath{\Omega} \cdot \B > 0$). We now explore how these results are 
modified when these vectors are antiparallel.

\begin{figure*}
\centerline{\epsfxsize=12cm \epsfbox{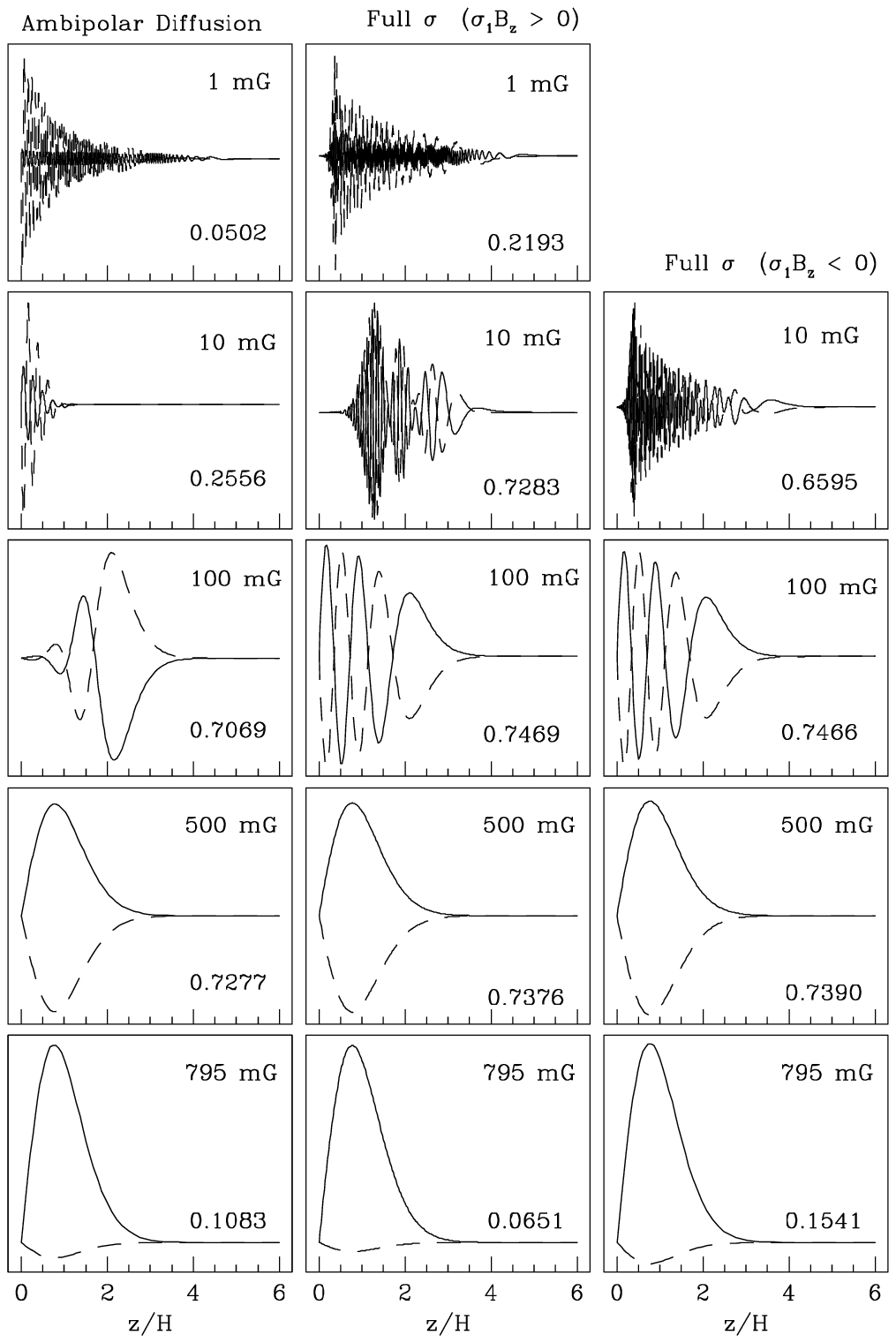}}\vskip 0cm
\caption{Comparison of the structure of the fastest growing modes for the 
fiducial model at $R = 5$ AU, for different configurations of the 
conductivity tensor. The left column corresponds to the ambipolar diffusion 
approximation. The middle and right columns show solutions obtained with a 
full conductivity tensor when 
the magnetic field and angular velocity 
vectors of the disc are parallel ($\sigma_1B_z > 0$) and antiparallel 
($\sigma_1B_z < 0$), respectively. Note that when $B \lesssim 10$ mG, full 
$\bmath{\sigma}$ solutions obtained with Hall conductivity terms of opposite 
signs are different, as this component dominates near the midplane in this 
region of parameter space. On the other hand, ambipolar diffusion perturbations
differ from full $\bmath{\sigma}$ ones for $B \lesssim 100$ mG, given that 
for these magnetic field strengths $\sigma_2$ dominates close to the midplane.
For stronger fields, the 
fluid is in nearly ideal-MHD conditions and perturbations in all three 
configurations of the conductivity tensor are alike.}
\label{fig:AUcond3}
\end{figure*}

As noted before, we explored in this case a reduced region of parameter 
space. In particular, solutions at $1$ AU could not be computed at all, 
as $\chi_o \ll 2$ at this radius, the limit below which all wavenumbers 
grow in this regime (W99, see also section \ref{subsec:test}). On the other 
hand, results at $10$ AU do not differ appreciably in this case from those 
obtained using a 
$\sigma_1B_z > 0$ conductivity, given the strong magnetic coupling throughout
the cross-section of the disc at this radial location. 
At $5$ AU, however, Hall diffusion is important for relatively weak fields 
($B \lesssim 10$ mG), and both sets of results should be different in this 
region of parameter space. Fig. \ref{fig:AUcond3} (middle and right columns) 
compares the structure of 
MRI 
unstable modes computed with a positive and negative Hall conductivity at 
this radius, as a function of the magnetic field strength. When 
$\sigma_1B_z < 0$, no results could be computed for 
$B < 8$ mG, as $\chi_o < 2$ in these cases. We find that, indeed, for 
$B = 10$ mG the most unstable mode computed with a negative Hall 
conductivity has a higher wavenumber, and a slower growth rate, than the 
corresponding mode with $\sigma_1B_z > 0$.
   
\subsubsection{More massive disc}
\label{subsubsec:surf_density}

\begin{figure*} 
\centerline{\epsfxsize=12cm \epsfbox{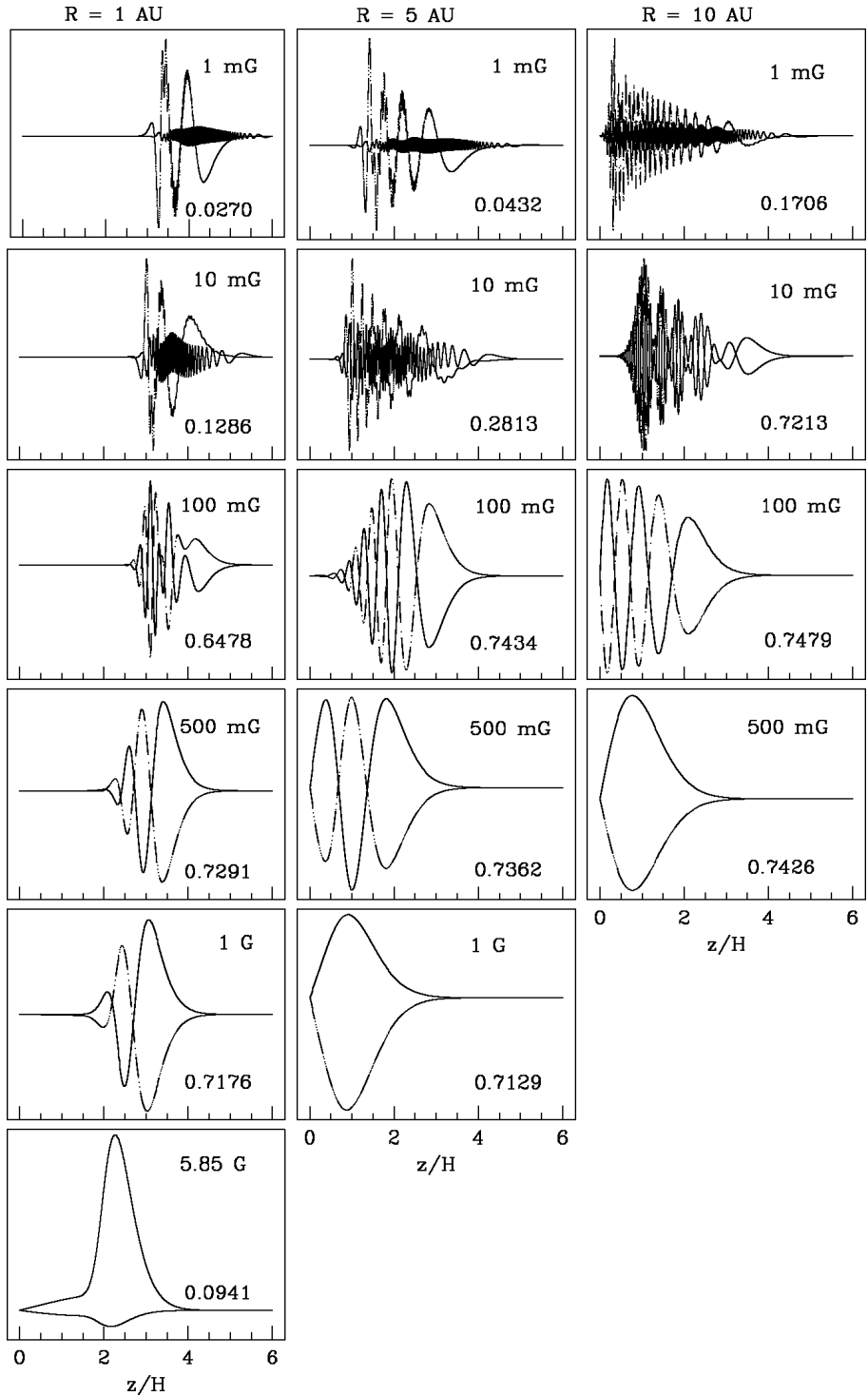}}\vskip 0cm
\caption{Comparison of the structure of the most unstable modes of the MRI for 
a more massive disc, incorporating cosmic ray ionisation. We 
present results at 
$R = 1$, $5$ and $10$  AU as a function of the strength of the magnetic 
field. The disc surface and mass density are $\Sigma_o^{'} = 10 \Sigma_o$ and 
$\rho_o^{'} = 10 \rho_o$. For simplicity, it was assumed that the 
temperature, 
sound speed and scale height are unchanged from those of the minimum-mass 
solar nebula model. 
Note that these modes have a more extended central dead zone, 
especially for $R = 1$ AU, in relation to the fiducial model. The kink in the 
perturbation at $R = 1$ AU for $B = 5.85$ G (leftmost 
column, bottom panel), is caused by the sudden change in the ionisation 
balance at the height where x-rays are not able to penetrate further within 
the disc.}
\label{fig:surfdensity}
\end{figure*}

To finalise the analysis of the structure of MRI perturbations, we describe 
now the properties of unstable modes in a more massive disc, as characterised
in section \ref{subsec:model}. Results are displayed in Fig. 
\ref{fig:surfdensity}. As expected, the larger column density causes unstable  
modes to have a more extended central dead zone in relation to results in the 
minimum-mass solar nebula model. This is particularly 
noticeable for $R = 1$ AU, where there is now a magnetically inactive zone 
for $B \lesssim 1$ G. X-rays are 
excluded from the midplane at this radius (they can only penetrate up to 
$z/H \sim 2.6$ for this disc model) and the cosmic ray ionisation rate at 
$z = 0$ is negligible. 
As a result, even for $B = 1$ G the magnetic 
coupling at the midplane is $\sim 10^{-4}$ and perturbations are damped at low
$z$. We observe here the same trends discussed in the analysis of the 
minimum-mass solar nebula model 
in relation to the structure of these modes. Ambipolar diffusion shapes the 
envelopes, particularly for strong fields, while Hall diffusion increases 
their wavenumber. Finally, note the kink on the 
amplitude of the $B = 5.85$ G mode, attributed to the sharp increase in the 
ionisation fraction at the height where x-ray ionisation becomes active. 

At $5$ AU, there is also a more extended dead zone up to $B \sim 100$ mG, in 
comparison with that of perturbations 
obtained using the minimum-mass solar nebula model. In that
case,  ideal-MHD conditions held throughout the cross-section of the disc for 
$B = 100$ mG, 
with $\chi_o \sim 20$ (see section \ref{subsubsec:numax}). On the contrary, 
in the present model, 
$\chi_o \sim 0.2$ for this magnetic field strength and ambipolar diffusion is 
locally dominant for $0.7 \lesssim z/H \lesssim 2$ (see 
Fig. \ref{fig:ionthick}, middle panel). As a result, the amplitude of this 
perturbation increases with $z$, a typical behaviour of ambipolar diffusion 
driven MRI, instead of being flat as before. 

Finally, for $R = 10$ AU, ideal-MHD conditions hold 
throughout the disc cross-section for $B \gtrsim 100$ mG (Fig. 
\ref{fig:ionthick}, bottom panel). Because of this, the envelopes are flat in 
this region of parameter space. On the other hand, when 
$B \sim 10$ mG, $\chi_o \sim 1.5$ and ambipolar diffusion dominates up to
$z/H \sim 2$. As a result, this mode exhibits a small central dead zone and 
the envelope peaks at an intermediate height. When $B \sim 1$ mG, $\chi_o 
\sim 0.05$ and Hall diffusion is dominant for $z/H \lesssim 2$. The central 
dead zone is much reduced, and this perturbation grows closer to the midplane 
despite the weak magnetic coupling. 

\subsection{Growth rate of the perturbations}
\label{subsec:growth}

\subsubsection{Fastest growing modes for different conductivity 
regimes}
\label{subsubsec:grow_cond}
  
\begin{figure} 
\centerline{\epsfxsize=8cm \epsfbox{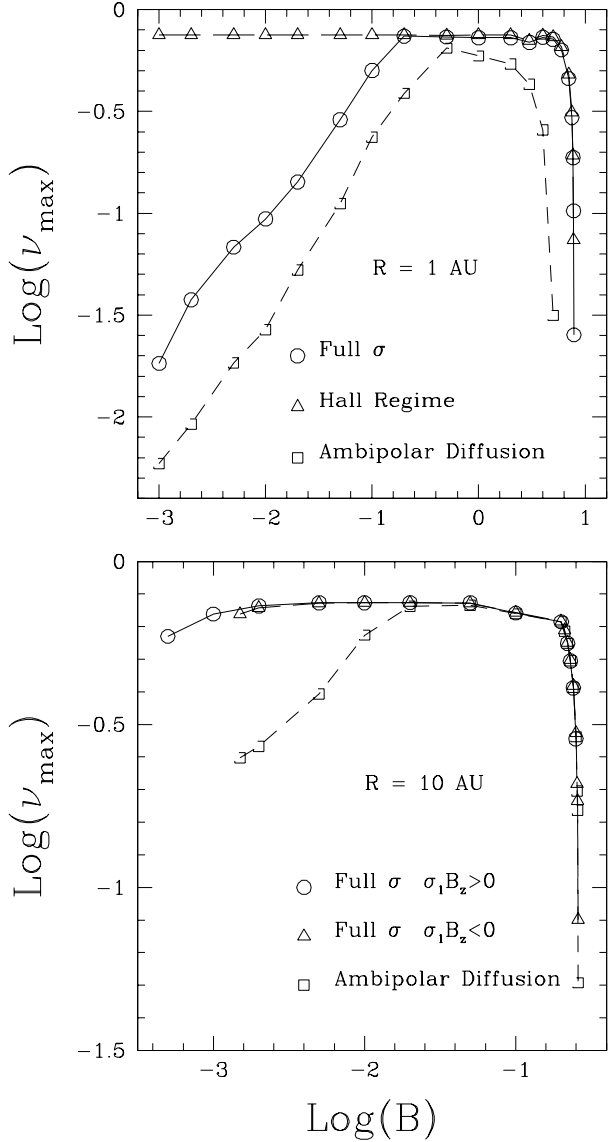}}\vskip 0cm
\caption{Growth rates of the fastest growing modes  as a function 
of the strength of the magnetic field for different configurations of the 
conductivity tensor. Results are presented at $R = 1$ AU (top panel) and $10$ 
AU (bottom panel) for the 
fiducial model. It is evident that at $1$ AU, when the magnetic field is weak 
($B \lesssim 200$ mG), perturbations are driven by ambipolar diffusion. 
As a result, $\nu_{max}$ increases with $B$. As the field gets 
even stronger, the maximum 
growth rate of full $\bmath{\sigma}$ and Hall limit perturbations are 
practically identical, signalling that Hall conductivity determines the growth 
of global unstable modes in this region of parameter space. Finally, 
$\nu_{max}$ decreases rapidly when $B$ is so strong that the wavelength of the 
fastest growing mode is $\sim$ the scaleheight of the disc.}
\label{fig:AUcond1}
\end{figure}

The dependency of the growth rate of the most unstable modes 
with the strength of the magnetic field is shown in Fig. 
\ref{fig:AUcond1} for different 
configurations of the conductivity tensor. Results correspond to the fiducial
model at $R = 1$ AU (top panel) and $10$ AU (bottom panel). 

The overall dependency of 
the $\nu_{max}$ vs. $B$ curve is typically as follows: $\nu_{max}$ 
initially increases with B with a $\sim$ power law dependency. It then 
levels at $ \sim 0.75$, the maximum growth rate for ideal-MHD 
perturbations in keplerian discs. Eventually, the maximum growth rate decays 
sharply at a 
characteristic magnetic field strength at which the perturbations' 
wavelength is $\sim H$, the scaleheight of the disc. Note that in the 
weak-field limit, the dependency of the $\nu_{max}$ vs. $B$ curve shows no 
evidence of a minimum field strength, below which modes do not grow. 
This characteristic 
shape of the $\nu_{max}$ vs. $B$ curve 
can be explained in terms of the dependency 
of the perturbations' growth rate with $B$ when different 
conductivity components are dominant at different heights. 

Examining Fig. \ref{fig:AUcond1} (top panel), it is clear that for weak fields 
($B \lesssim 200$ mG), perturbations 
computed with a full $\bmath{\sigma}$ have a $\nu_{max}$ vs. $B$ dependency 
similar to that of solutions obtained using the ambipolar 
diffusion approximation. This indicates that this feature is driven 
by ambipolar diffusion. In this limit,
the local maximum growth rate of MRI modes increases with the 
magnetic coupling (W99), which in turn, increases with $B$, except when 
$\beta_i$ and $\beta_e \gg 1$ (e.g. near the surface). 
As a result, $\nu_{max}$ should increase with $B$, as  
observed. 
On the contrary, when the magnetic field is stronger than $\sim 200$ mG, the 
growth rate of MRI modes obtained with 
a full conductivity tensor is practically identical to that of Hall limit 
modes. In this region of parameter space, the growth rate of global modes 
remains 
unchanged when the magnetic field gets stronger, as expected when Hall 
diffusion drives the growth of the instability. Eventually, $B$ becomes 
strong enough that the 
fastest growing mode becomes $\sim$ the scaleheight of the disc, and the 
growth rate rapidly declines. These results are in agreement with 
previous findings by W99 and SW03 .

Note also that the maximum growth rate of perturbations in the Hall limit 
does not change appreciably 
with $B$, until they are damped for a sufficiently strong field, as 
expected. In fact,  
Hall perturbations grow at $\sim$ the ideal-MHD rate even for $B \sim 1$ mG. 

Hall diffusion is locally dominant at $1$ AU in the lower sections of 
the disc for all magnetic field strengths for which unstable modes exist 
(Fig. \ref{fig:1AUion}, top panel). As a result, perturbations obtained with a 
full conductivity 
tensor grow significantly faster than modes found using the ambipolar 
diffusion approximation for all $B$.

For $R = 5$ AU, the Hall effect increases the growth rate of full 
$\bmath{\sigma}$ perturbations when $B \lesssim 10$ mG, the region of parameter
space where Hall diffusion is locally dominant near the midplane (compare the
growth rates in left and middle columns of fig. 
\ref{fig:AUcond3}. Moreover, full $\bmath{\sigma}$ modes computed with 
$\sigma_1B_z > 0$ (middle column) and $\sigma_1B_z < 0$ (right column) grow at 
different speeds and have adifferent structure, as expected in the region 
where Hall diffusion dominates. 

Finally, Fig. \ref{fig:AUcond1} (bottom panel) displays the growth rate of the 
most unstable modes at $10$ AU. Results were computed with a full conductivity 
tensor --but with opposite
signs of $\bmath{\Omega} \cdot \B$--, as well as using the 
ambipolar diffusion approximation. In this 
case, Hall diffusion is locally unimportant for all $B$ studied here, so the 
sign of $\sigma_1B_z$ does not affect the growth of the perturbations. On 
the other hand, at this radius, 
ambipolar diffusion is important for $B \lesssim 10$ mG. This slows the growth
of perturbations in the ambipolar diffusion limit when the magnetic field is 
weaker than this value.

\subsubsection{Fastest growing modes at diferent radii}
\label{subsubsec:nu_radii}

\begin{figure} 
\centerline{\epsfxsize=7.5cm \epsfbox{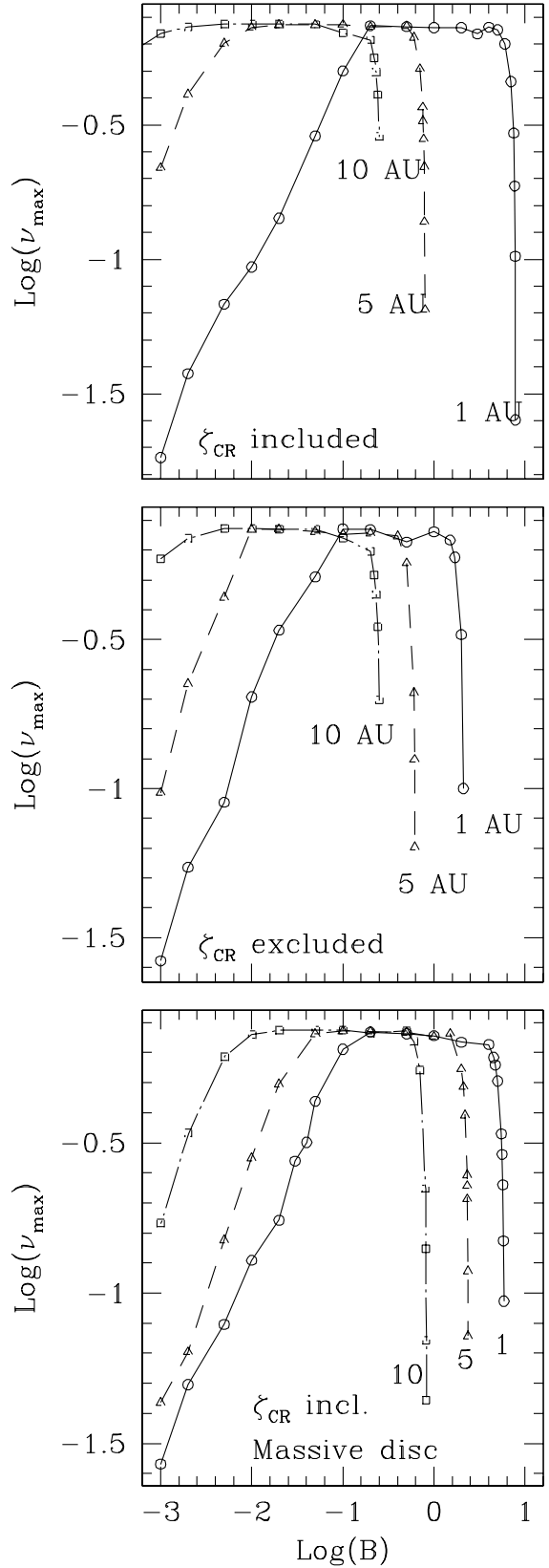}}\vskip 0cm
\caption{Growth rate of the most unstable modes of the MRI for $R = 1$, $5$ 
and $10$ AU as a function of the strength of the magnetic field. Top and 
middle panels show results obtained with the fiducial model, assuming 
cosmic rays are either included, or excluded, from the disc. Bottom panel 
displays solutions obtained with the more massive disc, including cosmic ray 
ionisation.}
\label{fig:growth}
\end{figure}

Finally, we compare the growth rate of the most unstable modes at different 
radii 
as a function of the strength of the magnetic field (Fig. \ref{fig:growth}). 
Three sets of results are displayed. The first two correspond to the 
minimum-mass solar nebula disc assuming cosmic 
rays either penetrate the disc (top panel) or are excluded from it 
(middle panel). The last set presents results for the more massive disc 
model (bottom panel), incorporating cosmic ray ionisation. In all three 
cases, MRI perturbations grow over a wide range of magnetic field strengths. 

Note that the maximum field strength 
for which unstable modes exist (within each panel), is weaker at larger radii. 
In ideal-MHD conditions, as well as when either the ambipolar 
diffusion or Hall ($\sigma_1B_z > 0$) conductivity regimes dominate, unstable 
MRI modes are damped 
when $v_A/c_s \sim 1$ (Balbus \& Hawley 1991), which corresponds to 
$\lambda \sim H$. In the Hall ($\sigma_1B_z < 0$) limit, unstable modes 
have been found for $v_A/c_s \sim 3$ (SW03). In any case, as both the gas 
density (eq. \ref{eq:rhoinitial}) and the sound speed (eq. 
\ref{eq:sound_speed}) decrease with radius, 
the ratio of the Alfv\'en to sound speed associated with a 
particular magnetic field strength increases with $R$, and as a result, the 
perturbations are damped at a weaker field for larger radii. 

The maximum magnetic field strengths for which MRI unstable modes grow, as 
well 
as the range for which $\nu_{max} \sim$ the ideal-MHD rate, are 
summarised in 
Table \ref{table:ionisation} for the radii and disc structures of interest 
here. Note that for the fiducial model, we obtained unstable modes at $1$ AU 
for $B \lesssim 8$ G. The growth rate is of the order of the 
ideal-MHD rate ($\nu = 0.75$) for $200$ mG $\lesssim B \lesssim 5$ G. When 
cosmic rays are assumed 
to be excluded from the disc (middle panel), unstable modes are obtained at 
$1$ AU only for 
$B \lesssim 2$ G. The corresponding range at $5$ AU is only slightly 
reduced and it is 
essentially the same as before at $10$ AU. This is consistent with our 
expectation that cosmic rays are a particularly important source of ionisation
at $1$ AU, where x-rays are excluded from the midplane. In 
this case, $\nu_{max} \sim 0.75$ at $1$ AU for $100$ mG 
$\lesssim B \lesssim 1$ G.

In the massive disc model (bottom panel of Fig. 
\ref{fig:growth}), the MRI is active at $1$ AU for $B \lesssim 6$ G and 
$\nu_{max} \sim 0.75$ for $200$ mG $< B < 500$ mG.
For $R = 5$ and $10$ AU, 
MRI unstable modes exist in this disc for stronger fields than in the 
minimum-mass solar nebula model. This can be explained by recalling 
that the larger mass and column density translates into a larger 
gas pressure and a stronger equipartition magnetic field 
strength at the midplane. MRI modes are, as a result, damped for a stronger
field than they are in the minimum-mass solar nebula model. On the contrary, 
for $R = 1$ AU, the range of 
magnetic field strengths over which perturbations grow does not change 
appreciably (see Table \ref{table:ionisation}).

\begin{table*}
\caption{Magnetic field strengths for which MRI perturbations grow 
at $1$, $5$ and $10$ AU for different disc models. For the minimum-mass solar 
nebula disc, two sets of results are shown, where cosmic rays are either 
present or excluded from the disc. For the massive disc model, only the former 
case is presented. Columns labeled `$B_{max}$'
list the maximum value of $B$ for which each disc supports unstable MRI 
modes. 
Similarly, the `$B(\nu \sim 0.75)$' columns specify the subset of these for 
which 
perturbations grow at nearly the ideal-MHD rate ($\nu \sim 0.75$).}

\begin{tabular}{lccccccccc}
\hline
\multicolumn{2}{c}{} &
\multicolumn{5}{c}{Minimum-mass solar nebula} & \multicolumn{1}{c}{} &
	\multicolumn{2}{c}{Massive Disc} \\
\multicolumn{2}{c}{} &
\multicolumn{2}{c}{$\zeta_{CR}$ included} & \multicolumn{1}{c}{} &
	\multicolumn{2}{c}{$\zeta_{CR}$ excluded} & \multicolumn{1}{c}{} & 
	\multicolumn{2}{c}{$\zeta_{CR}$ included} \\
\cline{3-4} \cline{6-7} \cline{9-10} 
Radius (AU) & & $B_{max}$ & $B(\nu \sim 0.75)$ & & $B_{max}$ & 
$B(\nu \sim 0.75)$ & & 	$B_{max}$ & $B(\nu \sim 0.75)$ \\
\hline
$1$ & & $7.85$ & $0.2$ - $5$ & & $2.10$ & $0.1$ - $1$ & &
 $5.85$ & $0.2$ - $0.5$ \\
$5$ & & $0.80$ & $0.02$ - $0.5$ & &$0.62$  & 
$0.01$ - $0.05$ & & $2.36$ & $0.05$ - $0.5$ \\
$10$ & & $0.25$ & $0.002$-$0.05$ & & $0.25$ & 
$0.005$ - $0.05$ & & $0.82$ & $0.01$ - $0.5$ \\
\hline
\end{tabular}
	\label{table:ionisation}
\end{table*}

\section{Discussion}
\label{sec:discussion}

In this paper we have explored the vertical structure and linear growth of 
MRI perturbations of an initially vertical magnetic field, using a 
realistic ionisation profile and assuming that 
ions and electrons are the sole charge carriers. This formulation is 
appropriate to model low conductivity discs (Gammie \& Menou 1998; Menou 
2000; Stone et al. 2000) at late stages of accretion, after dust grains 
have settled into a thin layer around the midplane ($\sim 10^5$ years, 
Nakagawa et. al. 1981, Dullemond \& Dominik 2004) and become 
dynamically uncoupled from the gas at higher $z$. Solutions were 
obtained at $1$, $5$ and $10$ AU from the central object as a function of the 
strength of the magnetic field for different configurations of the 
conductivity tensor, disc model and sources of ionisation. 

We have shown that the magnetic field is dynamically important in low 
conductivity accretion discs over 
a wide range of field strengths. An example of this activity is the 
generation and sustaining of MRI-unstable modes, which are needed to 
provide the angular momentum transport for accretion to proceed.  
The structure and growth rate of MRI perturbations are a function of the 
disc properties and the strength of the field. For a particular radius and 
disc model, they are a result of the competing action of different dominant
conductivity components at different heights. 

For $R \sim$ a few AU and relatively weak fields 
($B \lesssim 100$ mG at $1$ AU or 
$B \lesssim 1$ mG at $5$ AU), the fluid close to the midplane is in the 
Ohmic conductivity regime ($\sigma_{\parallel} \sim \sigma_2 \gg |\sigma_1|$). 
Conversely, for larger radii (or stronger fields) Hall conductivity
dominates even at $z = 0$ (Wardle, in preparation; see also section 
\ref{subsubsec:regimes}).  Ambipolar diffusion dominates above this region 
until $\chi \sim 10$ (W99). For still higher $z$, the 
magnetic coupling is so strong that ideal-MHD holds. The heights at which 
these 
transitions take place, for a particular disc model and radial position, are 
a function of the strength of the magnetic field. Note that even when 
$|\sigma_1| \ll \sigma_2$, Hall 
diffusion still dominates the structure and growth of MRI modes when 
$\chi \lesssim \chi_{crit} \equiv |\sigma_1|/\sigma_{\perp}$ (SW03).

At $1$ AU, for the fiducial model, unstable modes exist for 
$B \lesssim 8$ G. For a significant subset of these 
strengths ($200$ mG $\lesssim B \lesssim 5$ G), perturbations grow at about 
the ideal-MHD rate ($ = 0.75 \Omega$ 
in keplerian discs). At this radius, Hall diffusion is locally dominant near 
the midplane for all 
magnetic field strengths for which MRI unstable modes exist. As a result,
modes computed with a full conductivity
tensor have a higher wavenumber, and grow faster, than perturbations obtained
using the ambipolar diffusion approximation. On the other hand, ambipolar 
diffusion shapes the envelopes of these  
modes (especially for strong fields), causing them to peak 
at an intermediate height, instead of having
the characteristic flat envelope that modes in the Hall limit have for these 
field strengths. 

Cosmic rays are an important source of ionisation at $1$ AU, given that x-rays 
do not reach the midplane at this radius (see Fig. \ref{fig:ionbalance}). 
Consequently, if they are assumed to be excluded from the disc, 
the extent of the 
magnetically inactive `dead zone' about the midplane increases and unstable 
modes grow only for $B \lesssim 2.1$ G. Furthermore, under this assumption, 
perturbations obtained with a relatively strong initial magnetic field 
(e.g. $B \gtrsim 500$ mG), exhibit a kink in their amplitude. This 
feature is attributed to the sharp increase in the ionisation fraction at the
height where x-ray ionisation becomes active. Finally, for the massive disc 
model ($\Sigma_o^{'} = 10\Sigma_o$ and 
$\rho_o^{'} = 10\rho_o$, incorporating cosmic ray ionisation),
the central dead zone extends to a higher $z$ at $1$ AU, as the larger 
column density associated with this model causes the ionisation fraction at 
low $z$ to drop sharply (the disc is shielded from x-rays below 
$z/H \approx 2.6$ and even the cosmic ray ionisation rate at the midplane is 
negligible). MRI 
perturbations grow in this case for $B \lesssim 6$ G, a 
range not significantly different from that of results in the fiducial 
model. This is the case because the effective surface density is largely 
unchanged, given the shielding of the inner sections.

At $5$ AU, for the fiducial model, MRI modes grow for 
$B \lesssim 800$ mG. When $20$ mG $\lesssim B \lesssim 500$ mG, the 
growth rate is $\sim 0.75 \Omega$. At this radius, Hall diffusion increases 
the growth rate and wavenumber 
of unstable modes for $B \lesssim 100$ mG, but for stronger fields, 
perturbations 
obtained with different configurations of the conductivity tensor have 
similar structures and growth rates, a signal that ideal-MHD conditions hold 
(Fig. \ref{fig:AUcond3}). We conducted an analysis of the 
effect of the alignment of the magnetic field and the angular velocity 
vectors of the field. Results indicate that the sign of $\sigma_1 B_z$ 
is important for $B \lesssim 10$ mG, the range of magnetic field 
strengths for which Hall diffusion is locally dominant at low $z$. As in the 
$1$ AU case, excluding cosmic rays reduces the range of magnetic field 
strengths for which MRI modes grow (unstable modes are found for 
$B \lesssim 600$ mG in this case) and 
there is a kink in the amplitude of the perturbations for 
$B \gtrsim 100$ mG. Finally, when 
the surface density of the disc is increased, the central dead zone occupies a
larger cross-section, as expected, and unstable modes are found for stronger 
fields than in the fiducial model. This is a result of the stronger 
equipartition magnetic field strength for this disc model at this radial 
position.
 
At $10$ AU, the MRI is active for $B \lesssim 250$ mG and the 
growth rate is close to the ideal-MHD rate for $2$ mG 
$\lesssim B \lesssim 50$ mG. Furthermore, for $B \lesssim 10$ mG, 
perturbations obtained 
with a full conductivity tensor grow significantly faster than modes in the 
ambipolar diffusion limit, which reflects the contribution of the Hall effect 
(Fig. \ref{fig:AUcond1}, bottom panel). At this radius, when cosmic rays are 
excluded, the range of magnetic field strengths for which unstable modes exist 
is not affected, given that x-rays are able to penetrate to the midplane. 
In the more massive disc, the extent of 
the dead zone increases, especially for weak fields ($B \lesssim 10$ mG) and 
perturbations grow for stronger fields than in the 
minimum-mass solar nebula disc. 
 
In the present formulation, it was assumed that ions and electrons are the 
only charge carriers. As discussed earlier, this is valid in 
late evolutionary stages of accretion, when dust grains have sedimented enough
towards the midplane that they can be neglected when studying the dynamics of 
the gas at higher $z$. The timescale for this settling to occur is expected
to be affected by MHD turbulence. Although in quiescent discs dust grains may 
quickly settle
into a thin sub-layer about the midplane, the vertical stirring caused by MHD
turbulence could potentially transport them back to higher vertical 
locations, preventing them from settling below a certain height (Dullemond 
\& Dominik 
2004 and references therein). The efficiency of this process is dependent on 
the disc being able to support MHD turbulence in the vertical locations where 
dust grains are present. On the other hand, dust grains can reduce the 
abundance of free electrons, and the efficiency of MHD turbulence itself, 
by providing additional recombination pathways on their 
surfaces. To make things more complex, it is likely that the transition 
between vertical 
sections where dust grains are well mixed with the gas phase and those 
completely depleted of grains by settling, occur over a finite thickness 
(Dullemond \& Dominik 2004). Ultimately, the equilibrium
structure of accretion discs will reflect the complex interplay between all 
these processes. 

It is expected that the structure and growth of 
MRI unstable modes in such environments will be affected by the dependency 
of the ionisation balance with height in the presence of chemistry taking 
place on grain surfaces. The 
study of the properties of the MRI in a disc where dust dynamics and evolution
are determined consistently, and where Hall conductivity is taken into 
account, is essential to understand more fully the presence and efficiency of 
MRI-driven angular momentum transport in accretion discs. 
Nevertheless, from the results presented in this paper, it is clear that Hall 
diffusion is crucial for the realistic modelling 
of the magnetorotational instability in protostellar discs, particularly at a
distance of $\sim$ a few AU from the central object. More generally, Hall 
conductivity is an important factor when studying the magnetic activity of 
low conductivity discs at these radii.
\section{Summary}
\label{sec:summary}

We have presented in this paper the vertical structure and linear growth of 
the
magnetorotational instability (MRI) in weakly ionised, stratified accretion
discs, assuming an initially vertical magnetic field. Both the density and 
the conductivity are a function of height and the conductivity is treated as 
a tensor and obtained with a realistic ionisation profile. Two disc models 
were explored: The 
minimum-mass solar mebula disc (Hayashi 1981, Nakagawa \& Nakazawa 1985) and 
a more massive disc, with the mass and surface density increased by a factor 
of $10$.     
This formulation is appropriate for the study of weakly ionised
astrophysical discs, where the ideal-MHD approximation breaks down
(Gammie \& Menou 1998; Menou 2000; Stone et al. 2000). The ionisation sources
relevant here, outside the inner $0.1$ AU from the central object, are 
non-thermal: Cosmic rays, x-rays and radioactive decay. For the minimum-mass 
solar nebula model we compare solutions obtained including all three sources 
of ionisation with those arrived at assuming that cosmic rays are excluded 
from the disc by the protostar's winds. Recombination processes are taken to 
occur in the gas-phase only, which is consistent with the assumption 
that dust grains have settled into a thin layer about the midplane, and ions 
and electrons are the only charge carriers. Perturbations of interest have 
vertical wavevectors ($k = k_z$) only, which are the most unstable modes 
(when initiated from a vertically aligned 
magnetic field) in both the Hall and Ohmic regimes 
(Balbus \& Hawley 1991; Sano \& Miyama 1999). This is not 
necessarily the case in the ambipolar diffusion limit 
(Kunz \& Balbus 2003), where the fastest growing modes can have radial as well
as vertical wavenumbers. Under the adopted approximations, the properties of 
the MRI in the ambipolar diffusion and Ohmic conductivity limits are 
identical.

Three parameters were found to control the dynamics and 
evolution of the fluid: (\emph{i}) the local ratio of the Alfv\'en to sound 
speed ($v_A/c_s$); (\emph{ii}) The local coupling between ionised and neutral
components of the fluid ($\chi$), which relates the frequency at which
non-ideal effects are important with the dynamical (keplerian) frequency of
the disc; and (\emph{iii}) the ratio of the components of the conductivity 
tensor perpendicular to the magnetic field ($\sigma_1/ \sigma_2$), which 
characterises the conductivity regime of the fluid. These parameters were 
evaluated at $R = 1$, $5$ and $10$ AU for a range of magnetic field strengths.
The linearised system of ODE was integrated from the midplane to
the surface of the disc under appropriate boundary conditions and solutions 
were obtained for representative radial locations of the 
disc as a function of the magnetic field strength and for different 
configurations of the conductivity tensor.

The main results of this study are summarised below.

\begin{enumerate}
\item For the minimum-mass solar nebula model, incorporating cosmic ray 
ionisation (the fiducial model):

At $1$ AU, unstable MRI modes exist for $B \lesssim 8$ G. When $200$ mG
$\lesssim B \lesssim 5$ G, the most unstable modes grow at $\sim$ the 
ideal-MHD rate ($ = 0.75 \Omega$). 
Hall diffusion dominates the structure and growth rate of 
unstable modes for all magnetic field strengths for which they 
grow. For strong fields, ambipolar diffusion shapes the envelope of the 
perturbations, which peak at an intermediate height. Finally, at this radius, 
a magnetically dead zone (Gammie 1996, Wardle 1997) exists when $B < 1$ G. 
As expected, the vertical extent of this zone 
decreases when the magnetic field gets stronger and/or when a more massive 
disc model is used.

At $5$ AU, MRI modes grow for $B \lesssim 800$ mG and 
the growth rate is close to the ideal MHD rate for $20$ mG 
$\lesssim B \lesssim 500$ mG. 
Perturbations incorporating Hall conductivity have a higher wavenumber 
and grow faster than solutions in the ambipolar diffusion limit for 
$B \lesssim 100$ mG. Unstable modes grow even at the midplane for 
$B \geq 100$ mG but for weaker fields, a small dead region exists.
  
At $10$ AU, the MRI is active for $B \lesssim 250$ 
mG. The growth rate is close to the ideal-MHD rate for $2$ mG 
$\lesssim B \lesssim 50$ mG and when $B \lesssim 10$ mG, perturbations 
obtained with a full conductivity tensor grow significantly faster than 
modes in the ambipolar diffusion limit. Modes show only a very small dead 
region when $B \sim 1$ mG.

\item When the magnetic field is weak (e.g. $B \lesssim 200$ mG at $1$ AU), 
the maximum growth rate of unstable MRI modes ($\nu_{max}$) increases with 
the strength of the magnetic field, a feature driven by ambipolar diffusion.

\item When cosmic rays are assumed to be excluded from the disc, unstable 
modes at $1$ AU grow only for $B \lesssim 2.1$ G. Results at $5$ AU 
only change slightly, while solutions at $10$ AU are not affected at all, as 
expected.

\item For the massive disc model ($\Sigma_o^{'} = 10\Sigma_o$ and 
$\rho_o^{'} = 10\rho_o$, incorporating cosmic ray ionisation), MRI 
perturbations grow for stronger fields at $5$ and $10$ AU, in relation to the 
minimum-mass solar nebula model. Results at $1$ AU are unchanged, as in this 
case, the effective surface density is not significantly different.
\end{enumerate}

MRI perturbations grow in protostellar discs for a wide range of fluid 
conditions and magnetic field strengths. Hall diffusion largely determines 
the structure and growth rate of perturbations at radii of order of a few AU 
from the central protostar. This indicates that,
despite the low magnetic coupling, the magnetic field is dynamically 
important in low conductivity astrophysical discs and will impact the dynamics
and evolution of these discs.

\section{ACKNOWLEDGMENTS}
\label{sec:Acknowledgments}

This research has been supported by the Australian Research Council.


\bsp
\label{lastpage}
\end{document}